\documentclass[11pt]{article}
\usepackage{mymacros}

\title{Cosmological Entanglement Entropy from {the von Neumann Algebra of} Double-Scaled SYK\\ {\LARGE\it \& Its Connection with Krylov Complexity}}
\author[a]{Sergio E. Aguilar-Gutierrez\orcidlink{0000-0003-0308-0061}}
\affiliation[a]{Qubits and Spacetime Unit, Okinawa Institute of Science and Technology Graduate University,\footnote{\begin{CJK}{UTF8}{min}沖縄科学技術大学院大学\end{CJK}} 1919-1 Tancha, Onna, Okinawa 904 0495, Japan}
\emailAdd{sergio.ernesto.aguilar@gmail.com}
\abstract{We investigate entanglement entropy between {the pair of type II$_1$ algebras of} the double-scaled SYK (DSSYK) model {given a chord state}, its holographic interpretation {as generalized horizon entropy}; particularly {in the (anti-)de Sitter ((A)dS) space limits of the bulk dual}; and its connection with Krylov complexity. {{The density matrices in this formalism are operators in the algebras, which are specified by the choice of global state; and there exists a trace to evaluate their von Neumann entropy since the algebras are commutants of each other, which leads to a notion of algebraic entanglement entropy}}. We match it in {{triple-scaling}} limits to an area computed through a \emph{Ryu-Takayanagi formula} in (A)dS$_2$ space with entangling surfaces at {the asymptotic timelike or spacelike boundaries respectively}; providing a first-principles example of holographic entanglement entropy for (A)dS$_2$ space. This result reproduces the {Bekenstein-Hawking and }Gibbons-Hawking entropy {formulas} for specific entangling regions points, while it decreases for others. This construction does not display some of the puzzling features in dS holography. The entanglement entropy remains real-valued since the theory is unitary, and it depends on the Krylov {spread} complexity {of the Hartle-Hawking state. At last, we discuss higher dimensional extensions.}}
\begin{document}

\maketitle
\section{Introduction}\label{sec:intro}
\paragraph{Cosmological Entanglement Entropy from DSSYK}
The Ryu-Takayanagi (RT) formula \cite{Ryu:2006bv,Ryu:2006ef} and its extensions \cite{Hubeny:2007xt,Faulkner:2013ana,Engelhardt:2014gca} (see e.g.~\cite{Nishioka:2009un,Rangamani:2016dms,Chen:2019lcd,Harlow:2014yka} for reviews) tell us how minimal areas in the bulk (anchored to a boundary region) relate to entanglement entropy in the boundary theory in the anti-de Sitter (AdS)/conformal field theory (CFT) correspondence \cite{Maldacena:1997re}. This formula has been proven (in the classical limit of the bulk theory) through a replica trick in the gravitational path integral by \cite{Lewkowycz:2013nqa}, and it has been shown to obey all known entanglement entropy inequalities \cite{Hayden:2011ag} (including strong subadditivity in \cite{Headrick:2007km}). We are compelled to generalize the lessons from AdS/CFT holography to other spacetimes/boundary theories. We are particularly interested in answering the question
\begin{quote}
    \emph{What is holographic entanglement entropy beyond AdS/CFT?}
\end{quote}
This query is especially motivated by de Sitter (dS) space holography;\footnote{See initial developments in \cite{Strominger:2001pn,Witten:2001kn,Maldacena:2002vr}, and some recent approaches to holographic dS entropy in \cite{Lewkowycz:2019xse,Dong:2018cuv,Sato:2015tta,Hartle:2012tv,Hertog:2011ky,Susskind:2021esx,Shaghoulian:2022fop,Franken:2025lfy,Franken:2024ruw,Franken:2023pni,Bobev:2022lcc,Hikida:2021ese,Hikida:2022ltr,Nanda:2025tid,Narayan:2015vda,Narayan:2024fcp,Narayan:2022afv,Goswami:2021ksw,Fernandes:2019ige,Goswami:2024vfl,Narayan:2023zen,Narayan:2023ebn,Doi:2022iyj,Narayan:2020nsc,Narayan:2019pjl,Narayan:2017xca,Narayan:2015oka,Ruan:2025uhl,Coleman:2021nor,Silverstein:2022dfj,Batra:2024kjl,Kawamoto:2023nki,Aguilar-Gutierrez:2023hls,deBoer:2015kda}.} which will be the basis to answer the above question in dS$_2$ space by finding a geometric dual to entanglement entropy in an explicit boundary theory (see Sec.~\ref{sec:RT dS2} for details). One of the main goals in the dS holography program is to understand what are, if any, the microscopic degrees of freedom responsible for the Gibbons-Hawking (GH) entropy formula \cite{Gibbons:1977mu}; which is just a proposal to define the entropy associated with the cosmological horizon of dS space and it might not have an interpretation in terms of a statistical ensemble, although there is compelling evidence it indeed does \cite{Balasubramanian:2025hns} when there is an asymptotic time or null-like boundary. This could provide useful information about the dS space cosmological constant \cite{Galante:2023uyf,Harris:2023cow,Anninos:2012qw}. 
However, there are some difficulties in generalizing the RT formula for dS space. From the dS/CFT perspective \cite{Strominger:2001pn}, where the putative dual theory lives at $\mathcal{I}^\pm$, the most straightforward application of the RT formula (or its quantum extremal surface generalization \cite{Engelhardt:2014gca}) would imply that the boundary theory is not unitary \cite{Doi:2022iyj}.

On the other hand, lower-dimensional quantum gravity theories generically allow new opportunities to explicitly test concepts and observables  beyond previous frameworks; and in some cases, they might still retain similar features to higher dimensional analogues. For instance, recently, it has been argued that the black hole entropy in sine dilaton gravity \cite{Blommaert:2024whf} (which has also been developed in \cite{Blommaert:2023opb,Blommaert:2024ymv,Blommaert:2025avl,Blommaert:2025rgw,Blommaert:2025eps}; 
see also \cite{Bossi:2024ffa,Cui:2025sgy,Aguilar-Gutierrez:2025pqp,Aguilar-Gutierrez:2024oea,Heller:2024ldz,Heller:2025ddj}) does not satisfy a Bekenstein-Hawking (BH) entropy formula (reviewed in \cite{Ferrari:2025bxs}). 
This would then imply the RT formula might take a different form in more general spacetimes; or it might be a special feature in this model. A clear advantage of this bulk dual proposal is that one can learn about (A)dS$_2$ and flat space Jackiw–Teitelboim (JT) \cite{JACKIW1985343,TEITELBOIM198341} gravity (see \cite{Mertens:2022irh} for a review)
in limiting cases \cite{Blommaert:2024whf} from a concrete ultraviolet (UV) finite microscopic theory, the double-scaled Sachdev–Ye–Kitaev \cite{Sachdev:1992fk,kitaevTalks1,kitaevTalks2,kitaevTalks3} (DSSYK) model \cite{Berkooz:2018jqr,Berkooz:2018qkz,Erdos:2014zgc,Cotler:2016fpe} (see \cite{Berkooz:2024lgq} for a pedagogical review).\footnote{There are other proposals to holography of the DSSYK, which might be related one another \cite{Aguilar-Gutierrez:2025hty,Blommaert:2025eps,Aguilar-Gutierrez:2024oea}; most notably three-dimensional de Sitter (dS$_3$) space from an observer-centric \cite{Narovlansky:2023lfz,Verlinde:2024znh,Verlinde:2024zrh,Narovlansky:2025tpb,Aguilar-Gutierrez:2024nau,Blommaert:2025eps} and stretched horizon \cite{Susskind:2021esx,Susskind:2022bia,Susskind:2023hnj,Lin:2022nss,Rahman:2022jsf,Rahman:2023pgt,Rahman:2024iiu,Rahman:2024vyg,Sekino:2025bsc,Miyashita:2025rpt} perspectives; see also \cite{Milekhin:2023bjv,Okuyama:2025hsd,Yuan:2024utc,Gaiotto:2024kze,Tietto:2025oxn,Ahn:2025exp}.} 
For instance, it has been argued that one can deduce a precise notion of dS$_2$ holographic complexity \cite{Susskind:2014rva,Brown:2015bva,Brown:2015lvg,Couch:2016exn,Belin:2021bga,Belin:2022xmt} as Krylov complexity \cite{Balasubramanian:2022tpr,Parker:2018yvk} (see \cite{Nandy:2024htc,Rabinovici:2025otw,Baiguera:2025dkc} for reviews) in the DSSYK model. 

However, in a sharp contrast, entanglement entropy in the DSSYK model \cite{Goel:2023svz} and its place in the holographic dictionary beyond the AdS$_2$ limit \cite{Tang:2024xgg} remains much less developed than holographic complexity as Krylov complexity \cite{Ambrosini:2024sre,Rabinovici:2023yex,Heller:2024ldz,Aguilar-Gutierrez:2025mxf,Aguilar-Gutierrez:2025hty,Aguilar-Gutierrez:2025pqp,Aguilar-Gutierrez:2025sqh,Balasubramanian:2024lqk,Ambrosini:2025hvo,Fu:2025kkh}). So far, the literature has not addressed how {to evaluate the entanglement entropy in the double-scaled algebras of the DSSYK model}, which is one of the contributions in this work.\footnote{An alternative approach to the one in this work for defining entanglement entropy in this setting is to evaluate entanglement between different flavors of SYK models \cite{Iizuka:2024die} (which we comment about in Sec.~\ref{ssec:outlook bdry}); or considering entropy differences between states in a time-band \cite{Jensen:2024dnl} from an algebraic approach, which has been associated with entanglement in spatial subregions in \cite{Mertens:2025rpa}.}

\paragraph{{Algebraic Entanglement Entropy}}
States in gravity and gauge theories do not naturally factorize into subsystems. {In particular when the boundary theory has an infinite number of degrees of freedom, denoted by $N$, corresponding to quantum gravity at the disk level, there is a so-called factorisation problem \cite{Harlow:2018tqv}. This problem occurs when the bulk Hilbert space of a quantum gravity theory describing an asymptotically AdS two-sided black hole does not admit a tensor product structure expected from two CFT duals living at the asymptotic boundaries. This was originally investigated in JT gravity by \cite{Harlow:2018tqv}. It was argued that pure JT gravity (i.e.,~ without matter) is not holographic, and it was later found that by adding matter and finite $N$ corrections in the boundary theory (corresponding to higher genus topologies in the bulk path integral) one recovers the expected physical bulk Hilbert space factorization in JT gravity \cite{Boruch:2024kvv}, and general relativity \cite{Balasubramanian:2024yxk,Balasubramanian:2024rek,Balasubramanian:2025hns}. In the $N\rightarrow\infty$ case, while there is no Hilbert space factorization, there are algebraic approaches to define von Neumann entropy with a holographic interpretation \cite{Kolchmeyer:2023gwa,Penington:2023dql} once we account for matter contributions to the operator algebra. Otherwise,} one generically needs to apply an auxiliary map to recover a factorized Hilbert space (see e.g.~\cite{Lin:2017uzr,Lin:2018xkj,Jafferis:2019wkd,Mertens:2022ujr}), {{called the factorization map}}. However, so far, most work in this area has only been carried out in a gauge-{{fixed}} approaches, determined by a factorization map, and it generically requires postulating a set of (reasonable, but ultimately ad hoc) axioms that it should satisfy. A useful consistency check is that the resulting notion of bulk entanglement recovers the area term.

{We emphasize there is no need to formulate a factorization map to define entanglement entropy in the DSSYK model and study its holographic interpretation; given that it has a pair of type II operator algebras, called double-scaled algebras \cite{Lin:2022rbf,Xu:2024hoc}. However, entanglement entropy in the DSSYK model has not been carefully investigated at this point to the best of our knowledge, despite abundant literature on generalized entanglement entropy \cite{Kudler-Flam:2023qfl,Jensen:2023yxy,Chandrasekaran:2022cip,Chandrasekaran:2022eqq,Witten:2021unn,Witten:2023qsv,Witten:2023xze,Penington:2023dql} (among many others). There was an interesting work \cite{Tang:2024xgg} that motivates some of their formulas based on the double-scaled algebra \cite{Xu:2024hoc}. However, there was no justification for interpreting their expressions \cite{Tang:2024xgg} in terms of operator algebras; which is one of the goals in this work.}

{{Due to the lack of a concrete study rigorously defining von Neumann entropy based on the von Neumann algebra of the DSSYK model, and its bulk interpretation, we are prompted}} to examine the question
\begin{quote}
    \emph{What are the first principles to {{define entanglement entropy based on the double-scaled algebras of the DSSYK model? Does it have a bulk interpretation?}}}
\end{quote}

\paragraph{Aims and Results in This Work}{We will show one can define a density matrix associated to each of the relevant algebra of observables from a global pure state \cite{Witten:2021unn}. Given that the double-scaled algebras are commutants of each other \cite{Xu:2024hoc}, the von Neumann entropies of the density matrices associated to a given state in the chord Hilbert space for each algebra are equal to each other. Thus, we can define a notion of algebraic entanglement entropy,\footnote{{I thank Gonçalo Araújo-Regado for pointing out some of the results in this work could be thought of as entanglement entropy between algebras, which we formalize here.}} and discuss their bulk interpretation. In this work, we are particularly interested in the high energy limit of the DSSYK energy spectrum, which is associated to dS$_2$ space, as well as its AdS$_2$ black hole interpretation in the opposite low-energy limit.}

{{We implement the above approach to evaluate entanglement entropy and we }}study its consequences in this example of dS$_2$/CFT$_1$, consistent with the relation between sine dilaton gravity (which recovers dS JT gravity in a particular limit) and the DSSYK model \cite{Blommaert:2024whf,Blommaert:2024ymv,Cui:2025sgy,Blommaert:2025eps,Blommaert:2025rgw}. This leads to consistent expressions with other parts in the literature, and it generalizes them for arbitrary states in the chord Hilbert space. We interpret these expressions in terms of the dual bulk theory providing evidence for sine dilaton gravity \cite{Blommaert:2024ymv}, which is not an input in our calculations. In the bulk picture, our evaluation corresponds to the holographic entanglement entropy {where the entangling surfaces are located at $\mathcal{I}^\pm$}. The result indeed has the same structure as the generalized horizon entropy (i.e.~the quantum corrected Hubeny-Ragamani-Takayanagi \cite{Hubeny:2007xt,Faulkner:2013ana,Engelhardt:2014gca} formula) expected for a holographic bulk geometry dual to the DSSYK model. As we show in Sec.~\ref{ssec:examples entropy}, it also acquires a statistical interpretation once we derive its semiclassical limit in the Hartle-Hawking (HH) state \cite{hartle:1983ai}. 

Later, we match the boundary time-dependent entanglement entropy with an extremal area in dS$_2$ space. We recover for the first time a RT formula for dS$_2$/CFT$_1$ with a concrete boundary theory dual, which does not display some of the not-well understood features of dS/CFT in higher dimensional templates. To do this, we take a triple-scaling limit in the boundary theory {around the maximum of the energy spectrum}, which is defined in a way that the DSSYK Hamiltonian in this limit reproduces the generator of spatial displacements at $\mathcal{I}^\pm$ in dS$_2$ JT gravity, recently studied by \cite{Heller:2025ddj}. We evaluate the entanglement entropy in the DSSYK in this limit, and we compare it to the dilaton (i.e.~a codimension-two area) of dS JT gravity (see e.g. \cite{Svesko:2022txo} or App.~\ref{app:sine dilaton} for the definition of the bulk theory) at the RT surface in the bulk (which corresponds to a point in two-dimensional gravity). Manifestly, the boundary theory at $\mathcal{I}^\pm$ is \emph{still unitary}, in contrast to other approaches to dS/CFT (see e.g.~\cite{Vasiliev:1999ba,Vasiliev:1990en,Anninos:2011ui,Ng:2012xp,Anninos:2012ft,Hikida:2021ese,Hikida:2022ltr}). {We present analogous computation for the low-energy limit of the DSSYK energy spectrum and JT gravity, based on \cite{Tang:2024xgg}.}

We interpret the result in terms of sine dilaton gravity (more evidence for this relation in the dS JT gravity limit of sine dilaton gravity appears in \cite{Heller:2025ddj,Blommaert:2025rgw,Okuyama:2025hsd}) \emph{without} assuming its duality with the DSSYK model as input in our calculations. The entropy in the DSSYK model that we computed {in the upper edge of the energy spectrum} corresponds to holographic entanglement entropy between points at the asymptotic boundary of the effective AdS$_2$ black hole in sine dilaton gravity. Furthermore, there is \emph{no need to introduce auxiliary time-like boundaries} to define a thermal ensemble in the spacetime (e.g.~from the HH preparation of state); this is defined in the effective AdS geometry (see Sec. \ref{ssec:higher dS/CFT} for more details).

\paragraph{Connections with Krylov complexity}Our result shows a direct relation between the entanglement entropy from the boundary and bulk sides (where the entangling surface at $\mathcal{I}^\pm$ in dS$_2$) with the Krylov spread complexity (defined in \cite{Balasubramanian:2022tpr}) for the zero-particle HH state of the DSSYK model in the UV triple-scaling limit, corresponding to the geodesic length dressing the entangling surfaces at the asymptotic boundaries of (A)dS$_2$ space. In this state, the rate of growth of spread complexity takes similar form as in holographic complexity conjectures \cite{Brown:2015bva,Brown:2015lvg} (known as the Lloyd bound), which turns out to provide a lower bound on the growth of the exponent in holographic entanglement entropy.

All in all, our results show a rich connection between  entanglement entropy, spread complexity and QRFs. The factorization in the chord Hilbert space reveals new lessons for the holographic dictionary of the DSSYK model.

\paragraph{Plan of the manuscript}
In \textbf{Sec}.\,\ref{sec:Algebra Entropy}, {we define the entanglement entropy between the double-scaled algebras given a chord state in the DSSYK model,\footnote{{Note that this means von Neumann entropy for an operator in each of the algebras; rather than between all the operators of each algebra at once.}} while exemplifying the results} In \textbf{Sec}.\,\ref{sec:RT dS2}, we verify a RT formula in dS$_2$/CFT$_1$ by evaluating entanglement entropy in the boundary theory and matching it to an extremal area in the bulk. We discuss the differences with other settings, and we study connections between the results with Krylov spread complexity. We conclude in \textbf{Sec}.\,\ref{sec:disc}  with a summary{, extensions of our arguments for higher-dimensional (A)dS/CFT holography,} and some future directions.

We also provide complementary material in the appendices. In App.~\ref{app:notation} we summarize the notation used in this work. 
In App.~\ref{app:sine dilaton} we briefly review sine dilaton gravity and its connection with (dS) JT gravity for the bulk theory interpretation of some of our results; but we do not assume its duality with the DSSYK model for our derivations; our results are consistent with it instead. {In App.~\ref{app:WKB approx} we perform the Wentzel-Kramers-Brillouin (WKB) \cite{wentzel1926verallgemeinerung,kramers1926wellenmechanik,brillouin1926mecanique} approximation to evaluate the algebraic entanglement entropy defined in the main text in the triple-scaling limit.} App.~\ref{app:details evaluation RTdS2} contains further details about how to locate the RT surface in our calculation of {{holographic}} entanglement entropy in Sec.~\ref{sec:RT dS2}. In App.~\ref{app:relational EE} we provide more details about the relational interpretation of the  entanglement entropy from the bulk description. {At last, in App.~\ref{app:details speed limits} we study the Lanczos algorithm for Krylov spread and operator complexity to evaluate bounds on the rate of growth of Krylov complexity.}

\section{Entanglement Entropy from the Double-Scaled Algebra}\label{sec:Algebra Entropy}
In this section, after a very short review of the chord Hilbert space of the DSSYK model {we present the backbone of our work; where we define an entanglement entropy between the double-scaled algebras of the DSSYK model, and we study their properties.}

\paragraph{Outline}Before discussing the new results, in Sec.~\ref{ssec:rev chord space} we review the necessary concepts on chord Hilbert space to keep most of the discussion in the manuscript self-contained and to introduce notation. The rest of the manuscript contains the new material. {We then discuss about the double-scaled algebras and their entanglement entropy in Sec.~\ref{ssec:VNE}; afterward, we illustrate our findings in Sec.~\ref{ssec:examples entropy} through different examples.}

\subsection{Brief Review of Chord Hilbert Space}\label{ssec:rev chord space}
\begin{figure}
    \centering
    {\includegraphics[height=0.42\textwidth]{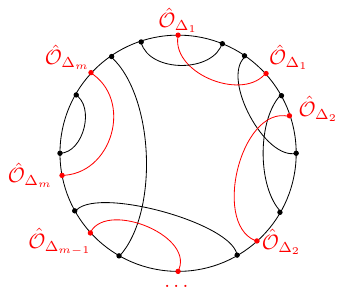}}
    \caption{Example of a chord diagram with $2m$ operator insertions $\hat{\mathcal{O}}_{\Delta_i}$.}
    \label{fig:many particle chord diagram}
\end{figure}

The chord Hilbert space of the DSSYK with $m$-particle insertions is represented by $\mathcal{H}_m$ \cite{Lin:2022rbf}
\begin{equation}\label{eq:states notation matter}
\mathcal{H}_m={\rm span}\qty{\ket{\tilde{\Delta};n_0,n_1,\dots,n_m}}~,
\end{equation}
where $\tilde{\Delta}=\qty{\Delta_1,\cdots,\Delta_m}$, represents a string of matter operator insertions, with  $\Delta_{i}$ being the conformal dimension of the matter chord operator for
\begin{equation}\label{eq:reference pets}
\hat{\mathcal{O}}_{\tilde{\Delta}}=\qty{\hat{\mathcal{O}}_{\Delta_{1}},\dots,~\hat{\mathcal{O}}_{\Delta_{m}}}~,  \quad m\in\mathbb{Z}_{>0}~;
\end{equation}
while $n_0$ is the number of DSSYK chords (called H-chords) to the left of all matter chords, $n_1$ the number between the first two particles; all the way up to the number of chords between all the $m$ particles. We also denote $\ket{\Omega}$ as the zero chord number associated with the maximal entropy state in the DSSYK, \cite{Lin:2022rbf}, {which is manifest from the algebraic perspective, as seen in Sec.~\ref{ssec:examples entropy}}. We illustrate this in Fig.~\ref{fig:many particle chord diagram}.
The evolution of states in $\mathcal{H}_m$ is generated by the DSSYK two-sided Hamiltonian, which was originally constructed in \cite{Lin:2022rbf} in terms of creation ($\hat{a}_i^\dagger$) and annihilation ($\hat{\alpha}_i$) operators as
\begin{subequations}\label{eq:two_sided_DSSYK_H}
    \begin{align}
\hH_L&=-\frac{J}{\sqrt{\lambda}}\qty(\hat{a}^\dagger_0+\sum_{i=0}^m\hat{\alpha}_i\sqrt{[\hat{n}_i]_q}q^{\hat{n}_i^<})\quad\text{where}\quad\hat{n}_i^<=\sum_{j=0}^{i-1}\qty(\hat{n}_j+\Delta_{j+1})~,\label{eq:HLmultiple}\\
\hH_R&=-\frac{J}{\sqrt{\lambda}}\qty(\hat{a}^\dagger_m+\sum_{i=0}^m\hat{\alpha}_i\sqrt{[\hat{n}_i]_q}q^{\hat{n}_i^>})\quad\text{where}\quad\hat{n}_i^>=\sum_{j={i+1}}^{m}\qty(\hat{n}_j+\Delta_{j})~,\label{eq:HRmultiple}
\end{align}
\end{subequations}
where $J$ is a coupling constant, and $q=\rme^{-\lambda}\in[0,1)$ is a parameter of the model; while
\begin{subequations}\label{eq:Fock Hm}
    \begin{align}\label{eq:Fock Hm 1}
    \hat{a}^\dagger_{i}\ket{\Delta_1,\dots,\Delta_m;n_0,\dots n_i,\dots, n_m}&=\sqrt{[n_i+1]_q}\ket{\Delta_1,\dots,\Delta_m;n_0,\dots, n_i+1,\dots n_m}\\\label{eq:Fock Hm 2}
    \hat{\alpha}_{i}\ket{\Delta_1,\dots,\Delta_m;n_0,\dots n_i,\dots, n_m}&=\ket{\Delta_1,\dots,\Delta_m;n_0,\dots, n_i-1,\dots n_m}~.
\end{align}
\end{subequations}
We will often refer to the two-sided HH state in $\mathcal{H}_1$ as
\begin{equation}\label{eq:HH state tL tR}
\ket{\Psi_{{\Delta}}(\tau_L,\tau_R)}=\rme^{-\tau_L\hH_L-\tau_R\hH_R}\hat{\mathcal{O}}_{\Delta}\ket{\Omega}=\rme^{-\tau_L\hH_L-\tau_R\hH_R}\ket{\Delta;0,0}~.
\end{equation}
In the following, the analytic continuation for the HH state is denoted $\tau_{L/R}=\frac{\beta_{L/R}}{2}+\rmi t_{L/R}$, where $\beta_{L/R}$ are two-sided inverse temperatures, $t_{L/R}$ similarly denote real time parameters. Thus this can be used to define partition functions in the one-particle space
\begin{equation}\label{eq:partition 1 particle}
    Z_\Delta(\beta_L,\beta_R):=\bra{\Psi_\Delta(\tau_L,\tau_R)}\ket{\Psi_\Delta(\tau_L,\tau_R)}=\bra{\Delta;0,0}\rme^{-\beta_L\hH_L-\beta_R\hH_R}\ket{\Delta;0,0}~.
\end{equation}
{Next, the zero-particle Hamiltonian is a limiting case of $\hH_L=\hH_R$ when $\Delta=0$, which can be expressed as
\begin{equation}\label{eq:H zero particle}
    \hH=\frac{J}{\sqrt{\lambda}}\qty(\ha+\ha^\dagger)~,
\end{equation}
which acts on the zero-particle chord number basis $\qty{\ket{n}}$ as
\begin{equation}\label{eq:q-oscillator algebra}
    \ha^\dagger\ket{n}=\sqrt{[n+1]_q}\ket{n+1}~,\quad\ha\ket{n}=\sqrt{[n]_q}\ket{n-1}~,\quad\hn\ket{n}=n\ket{n}~,
\end{equation}
with $[n]_q=(1-q^n)/(1-q)$, and $\bra{n}\ket{m}=\delta_{nm}$. This basis can be constructed by defining a normal ordering operator acting on a dimensionless Hamiltonian
\begin{equation}\label{eq:def h}
    \hat{h}\equiv(\sqrt{\lambda}/J)\hH~,
\end{equation}
so that $\ket{n}=:\hat{h}^n:\ket{\Omega}$, where
\begin{equation}\label{eq:normal order}
\sqrt{[n+1]_q}:\hat{h}^{n+1}:\equiv\hat{h}:\hat{h}^n:-\sqrt{[n]_q}:\hat{h}^{n-1}:~.
\end{equation}
Note that $:\hat{h}^n:$ is simply a n-degree polynomial in $\hat{h}$ by applying the recurrence relation above.}

The zero-particle space HH state counterpart of \eqref{eq:HH state tL tR} is denoted
\begin{equation}\label{eq:HH state}
    \ket{\Psi(\tau)}=\rme^{-\hH\tau}\ket{\Omega}~,
\end{equation}
where $\tau=\frac{\beta}{2}+\rmi t$ with $\beta_{L}=\beta_{R}=\beta$ in \eqref{eq:HH state tL tR}, $t_L=t_R=t$. This can be used to evaluate the zero-particle partition function $Z(\beta)=\bra{\Psi(\tau)}\ket{\Psi(\tau)}$.

We also describe energy states where $\hH_{L/R}$ (or $\hH$) become diagonal:
\begin{subequations}
    \begin{align}
        \label{eq:energy basis}
    \hH_{L/R}\ket{\Delta;\theta_L,\theta_R}&=E(\theta_{L/R})\ket{\Delta;\theta_L,\theta_R}~,\quad
\hH\ket{\theta}=E(\theta)\ket{\theta}~,\\
    E(\theta)&=-\frac{2J}{\sqrt{\lambda(1-q)}}\cos\theta~.\label{eq:energy}
    \end{align}
\end{subequations}
The energy basis is normalized such that:
\begin{align}
    &\mathbb{1}=\int_{0}^{\pi}\rmd\theta~\mu(\theta)\ket{\theta}\bra{\theta}~,\quad \mu(\theta):=\frac{1}{2\pi}(q,~\rme^{\pm 2 \rmi \theta};q)_\infty~,\quad \label{eq:norm theta}\\
{\rm where}~~&(a;~q)_n:=\prod_{k=0}^{n-1}(1-aq^k)~,\quad
    (a_0,\dots, a_N;q)_n:=\prod_{i=1}^N(a_i;~q)_n~.\\
\label{eq:identity theta}
&(x^{\pm a_1\pm a_2};q)_n:=(x^{a_1+ a_2};q)_n(x^{- a_1+a_2};q)_n(x^{-a_1+ a_2};q)_n(x^{-a_1-a_2};q)_n~,
\end{align}
and the inner product between energy and chord zero-particle basis elements is given in terms of q-Hermite polynomials ($H_n(x|q)$) \cite{Berkooz:2018jqr}:
\begin{equation}\label{eq:H_n def}
    \bra{\theta}\ket{n}={\frac{H_n(\cos\theta|q)}{\sqrt{(q;q)_n}}}~,\quad H_n(\cos\theta|q):=\sum_{k=0}^n\frac{(q;~q)_{n}~\rme^{\rmi(n-2k)\theta}}{(q;~q)_{n-k}(q;~q)_{k}}~.
\end{equation}
{Similar to \eqref{eq:norm theta}, the resolution of the identity in the one-particle space is given by \cite{Xu:2024gfm}
\begin{equation}\label{eq:norm theta 1p}
    \mathbb{1}=\int\rmd\theta_L\rmd\theta_R~\mu(\theta_L)\mu(\theta_R)\hat{\mathcal{F}}_\Delta^\dagger\qty(\ket{\theta_L}\otimes\ket{\theta_R})\qty(\bra{\theta_L}\otimes\bra{\theta_R})\hat{\mathcal{F}}_\Delta~,
\end{equation}
with the isometric factorization map \cite{Xu:2024hoc,Xu:2024gfm,Okuyama:2024yya,Okuyama:2024gsn} $\hat{\mathcal{F}}_\Delta:~\mathcal{H}_1\rightarrow\mathcal{H}_0\otimes\mathcal{H}_0$ which can be defined by its action on the energy basis as
\begin{equation}\label{eq:isometric factorization}
    \hat{\mathcal{F}}_\Delta\ket{\Delta;\theta_L,\theta_R}=\sqrt{\bra{\theta_L}q^{\Delta\hat{n}}\ket{\theta_R}}\ket{\theta_L}\otimes\ket{\theta_R}~.
\end{equation}
In the following, we employ the above concepts to define and evaluate the entanglement entropy associated to operators acting on the chord Hilbert space.}

\subsection{Algebraic Entanglement Entropy}\label{ssec:VNE}
{
In this section, we investigate the entanglement entropy associated with the double-scaled algebras of the DSSYK model for a given state in the chord Hilbert space. Generalized entropies in quantum gravity have received much recent attention \cite{Witten:2021unn,Chandrasekaran:2022cip,Chandrasekaran:2022eqq,Jensen:2023yxy,Kudler-Flam:2023qfl} among others. Great reviews on this subject can be found in \cite{Witten:2018zxz,Sorce:2023fdx,Casini:2022rlv,Liu:2025krl}. Original work connecting this formalism to holography can be explored in \cite{Papadodimas:2013jku,Jefferson:2018ksk,Leutheusser:2021frk,Leutheusser:2022bgi,Witten:2021unn}.}

{Chord operators acting on the DSSYK model form a pair of double-scaled algebras \cite{Lin:2022rbf,Xu:2024hoc}
\begin{equation}\label{eq:double scaled algebra}
\mathcal{A}_{L/R}=\qty{\hH_{L/R},\hat{\mathcal{O}}_\Delta^{{L/R}}}~,    
\end{equation}
which are commutants of each other i.e.
\begin{equation}
    [\hat{w}_L,\,\hat{w}_R]=0~,\quad\forall\hat{w}_{L/R}\in\mathcal{A}_{L/R}~.
\end{equation}
The states in the chord Hilbert space with all possible matter insertions
\begin{equation}\label{eq:H ffull}
    \mathcal{H}_{\rm chord}=\bigoplus_{m=0}^\infty\mathcal{H}_m~,
\end{equation}
are generated by the Gelfand-Naimark-Segal \cite{gelfand1943imbedding,Segal1947IrreducibleRO} (GNS) construction acting on the cyclic separating state, which is the zero chord state $\ket{\Omega}$ \cite{Xu:2024hoc}. It was shown in \cite{Xu:2024hoc} that the notion of trace in the left/right algebras is,\footnote{{Note that the algebra trace is defined to recover finite expectation values for trace class operators in the algebra, while the Hilbert space trace would lead to a generically diverging answer \cite{Chandrasekaran:2022cip,Witten:2021unn}.}}
\begin{equation}\label{eq:algebra trace}
    \Tr_{{L/R}}[\hat{w}_{L/R}]=\bra{\Omega}\hat{w}_{L/R}\ket{\Omega}~,
\end{equation}
where $\Tr_{{L/R}}(\hat{w}_{L/R}^\dagger \hat{w}_{L/R})\geq0$. Note that the definition of the trace is \emph{unique up to a non-zero constant rescaling,} corresponding to a different normalization for $\ket{\Omega}$.}

{We can also define a density matrix $\hrho_{L/R}\in\mathcal{A}_{L/R}$ associated to a state $\ket{\Psi}\in\mathcal{H}_{\rm full}$ \eqref{eq:H ffull}, namely:
\begin{equation}\label{eq:rho from state definition}
    \bra{\Psi}\hat{w}_{L/R}\ket{\Psi}=\bra{\Omega}\hat{\rho}_{L/R}\hat{w}_{L/R}\ket{\Omega}~,\quad\forall\hat{w}_{L/R}\in\mathcal{A}_{L/R}~.
\end{equation}
Suppressing indices, the von Neumann entropy of the density matrices $\hat{\rho}\in\mathcal{A}$, is defined using the algebraic trace \eqref{eq:algebra trace} as
\begin{equation}\label{eq:algebraic entropy}
\begin{aligned}
        S(\hrho)&\equiv\log\Tr\hrho-\frac{\Tr(\hrho\log\hrho)}{\Tr\hrho}\\
        &=\log\bra{\Omega}\hrho\ket{\Omega}-\frac{\bra{\Omega}\hrho\log\hrho\ket{\Omega}}{\bra{\Omega}\hrho\ket{\Omega}}~,
\end{aligned}
\end{equation}
where $\hrho$ is not necessarily normalized, one can choose to normalize it ${\bra{\Omega}\hrho\ket{\Omega}}=1$ to simplify the above expression as $\Tr(\hrho\log\hrho)$.
\\
Given that $\ket{\Psi}$ \eqref{eq:rho from state definition} is a pure global state, it follows that $S(\hrho_L)=S(\hrho_R)$. This means that \eqref{eq:algebraic entropy} can be used as a notion of entanglement entropy between the algebras $\mathcal{A}_L$ and $\mathcal{A}_R$ for a given chord state. For this reason, we will denote \eqref{eq:algebraic entropy} as \emph{algebraic entanglement entropy}. The arguments above are model-independent; \eqref{eq:algebraic entropy} is defined from $\hrho_{L/R}\in\mathcal{A}_{L/R}$ with $\mathcal{A}_{L}=(\mathcal{A}_{R})'$ which is obtained from a pure global state $\ket{\Psi}$ through \eqref{eq:rho from state definition}.}

{A summary of the procedure is shown in Tab.~\ref{tab:procedure}.
\begin{table}[t]
    \centering
    \begin{tabular}{cc}\toprule[2.0pt]
    \textbf{Step}&\textbf{Relation}\\\midrule[2.0pt]
        Choose a state in the chord Hilbert space $\ket{\Psi}\in\mathcal{H}_{\rm full}$ & \eqref{eq:H ffull}\\\hline
        Find the corresponding density matrices $\hrho_{L/R}\in\mathcal{A}_{L/R}$ & \eqref{eq:rho from state definition}\\\hline
        Evaluate their von Neumann entropy $S(\hrho_{L})=S(\hrho_{R})$ & \eqref{eq:algebra Renyi entropy}\\\bottomrule[2.0pt]
    \end{tabular}
    \caption{{A procedure to evaluate the entanglement entropy between the double-scaled algebras $\mathcal{A}_{L/R}$ given a chord state $\ket{\Psi}$.}}
    \label{tab:procedure}
\end{table}
From now on, we will suppress indices $L/R$ unless explicitly stated. Similarly, one can define the Rényi entropy for $\hrho$ using the algebra trace as,
\begin{equation}\label{eq:algebra Renyi entropy}
    S_\alpha\equiv\frac{1}{1-\alpha}\log\frac{\Tr\hrho^\alpha}{(\Tr\hrho)^\alpha}~.
\end{equation}
To be more specific about $\hrho$; any state $\ket{\Psi}\in\mathcal{H}_{m}$ can be built from the cyclic separating state $\ket{\Omega}$ through the GNS construction by an iterative application of the operators in either of the algebras $\mathcal{A}_{L/R}$. Let us then denote the state \begin{equation}
    \ket{\Psi}=f(\hH_{L/R},\hat{\mathcal{O}}^{L/R}_\Delta)\ket{\Omega}\in\mathcal{H}_{\rm chord}~,
\end{equation}
where $f$ is an analytic function, which admits a power series expansion in its arguments $\hH_{L/R}$ and $\hat{\mathcal{O}}^{L/R}_\Delta$. Then, we can identify the corresponding reduced density matrix from \eqref{eq:rho from state definition} as
\begin{equation}
    \hrho_{L/R}=f(\hH_{L/R},\hat{\mathcal{O}}^{L/R}_\Delta)f(\hH_{L/R},\hat{\mathcal{O}}^{L/R}_\Delta)^\dagger~,
\end{equation}
and its von Neumann entropy is just \eqref{eq:algebraic entropy}. This can be simplified in the energy basis using the resolution of the identity in the one-particle \eqref{eq:norm theta 1p} and zero-particle chord space \eqref{eq:norm theta}. We illustrate this below.}

\paragraph{Special case: No Operator Insertion}{
Consider an arbitrary state in the zero-particle chord space, which can be written as \cite{Xu:2024hoc} $\ket{\Psi}=f_e(\hH)\ket{\Omega}$ with $f_e$ being an analytic real-valued function. The corresponding density matrix \eqref{eq:rho from state definition} is $\hat{\rho}=f_e(\hH)f_e(\hH)^\dagger=(f_e(\hH))^2$.
It follows from \eqref{eq:norm theta} that the entanglement entropy \eqref{eq:algebraic entropy} takes the form:
\begin{equation}\label{eq:EE new}
\begin{aligned}
S&=-\int_0^\pi\rmd\theta\mu(\theta)\frac{({f_e(E(\theta))})^2}{\bra{\Psi}\ket{\Psi}}\log(\frac{({f_e(E(\theta))})^2}{\bra{\Psi}\ket{\Psi}})~.
\end{aligned}
\end{equation}
We may express \eqref{eq:EE new} in terms of a probability distribution as,
\begin{equation}\label{eq:entropy probability dist}
S=\int_0^\pi\rmd\theta\qty(-p(\theta)\log p(\theta)+p(\theta)\log\mu(\theta))~,
\end{equation}
where
\begin{equation}
p(\theta)\equiv \mu(\theta)\frac{({f_e(E(\theta))})^2}{\bra{\Psi}\ket{\Psi}}~,\quad\int_0^\pi\rmd\theta~ p(\theta)=1~.
\end{equation}
Note that the normalization factor $\bra{\Psi}\ket{\Psi}$ guarantees that the probability distribution $p(\theta)$ is normalized to one. This is consistent with entanglement entropy from density matrix decompositions in App.~I of \cite{Araujo-Regado:2025ejs}.}

\paragraph{Comparison with Other Works}
{First, to compare the entanglement entropy formula \eqref{eq:entropy probability dist} with JT gravity, we consider the infrared (IR) triple-scaling limit,\footnote{{See Sec.~\ref{sec:RT dS2} for details about our notation of IR and UV triple-scaling limits.}} where:
\begin{equation}\label{eq:triple scaling}
    \theta\rightarrow\lambda s~,\quad \mu(\theta)\rightarrow\rho(s)\equiv 2s\sinh(2\pi s)~,\quad s\geq0~,
\end{equation}
so that \eqref{eq:entropy probability dist} becomes equivalent to the quantum RT formula in JT gravity \cite{Jafferis:2019wkd} (8.107).
\\
Meanwhile, \cite{Tang:2024xgg} postulated a similar formula to \eqref{eq:entropy probability dist} for fixed chord number states. In contrast, this manuscript justifies that \eqref{eq:entropy probability dist} corresponds to algebraic entanglement entropy for a specific chord state. Importantly, \cite{Tang:2024xgg} found that in the triple-scaling limit the evaluation of the entanglement entropy for fixed chord number states ($\ket{n}$) reproduces the RT formula in JT gravity when the entangling regions are points at the asymptotic AdS$_2$ boundaries. This is a special case of our results, that we extend also to the dS$_2$ JT gravity in Sec.~\ref{sec:RT dS2}. Later, we relate the results with sine dilaton gravity, without assuming it as an input in our calculations. While one could assume a duality between the DSSYK model with sine dilaton gravity; the manuscript takes a more agnostic perspective; the results provide evidence for the sine dilaton interpretation without assuming a specific bulk dual in our computations in Sec.~\ref{sec:RT dS2}.\footnote{{Similarly, one does not need to assume a specific bulk dual to associate the canonically quantized Arnowitt–Deser–Misner \cite{Arnowitt:1959eec,Arnowitt:1960es,Arnowitt:1960zza,Arnowitt:1960zzb,Arnowitt:1960zzc,Arnowitt:1961zz,Arnowitt:1961zza} (ADM) Hamiltonian of JT gravity at the disk level \cite{Harlow:2018tqv} with the IR triple-scaling limit of the chord Hamiltonian \cite{Lin:2022rbf}; instead, it is an observation that they are isomorphic to each other; and thus, that they provide a concrete realization of AdS$_2$/CFT$_1$. Similarly, we propose a UV triple-scaling limit of the DSSYK chord Hamiltonian, which we find that it matches the generator of spatial translations along $\mathcal{I}^\pm$ in dS$_2$ JT gravity found by \cite{Heller:2025ddj}.}}
\\
At last, our formulation for entanglement entropy between the double-scaled algebras given (\ref{eq:entropy probability dist}) also has the same form as a conjectured entanglement entropy formula in \cite{Belaey:2025ijg} (5.1) based on the quantum group structure of DSSYK. There is no obvious connection to the quantum group of the DSSYK model in the evaluation leading to the formula. I thank Thomas Tappeiner for pointing out. We discuss in more detail in Sec.~\ref{sec:disc}.}

\subsection{Black Hole Entropy \& Partially-Entangled Thermal States}\label{ssec:examples entropy}
{
To illustrate the above results, we set up useful examples for discussing the bulk dual of the DSSYK model, which we elaborate in Sec.~\ref{sec:RT dS2}.}

\paragraph{Black Hole Entropy Without Matter}{We consider $\ket{\Psi}=\rme^{-\tau\hH}\ket{\Omega}$, where $\tau=\rmi t+\frac{\beta}{2}$ with $t\in\mathbb{R}$ and $\beta\in\mathbb{R}$. The density matrix is $\hat{\rho}=\rme^{-\beta\hH}$,  and the entanglement entropy \eqref{eq:entropy probability dist} becomes
\begin{equation}
\begin{aligned}\label{eq:entropy}
    S^{(\rm thermal)}=&\int_0^\pi\rmd\theta~\mu(\theta)\frac{\rme^{-\beta E(\theta)}}{Z(\beta)}\qty(\beta E(\theta)+\log Z(\beta))\\
    =&\qty(1-\beta\pdv{\beta})\log Z(\beta)~.
\end{aligned}
\end{equation}
where we denote $Z(\beta)\equiv \bra{\Omega}\rme^{-\beta\hH}\ket{\Omega}$. This is indeed the statistical entropy for a general quantum system in the canonical ensemble seen as entanglement entropy between the double-scaled algebras.
\\
In particular, in the semiclassical limit one finds \cite{Goel:2023svz}
\begin{equation}
S^{(\rm thermal)}\eqlambda \eval{S_0-\frac{1}{\lambda}(\theta-\pi/2)^2}_{\beta=\frac{\pi-2\theta}{J\sin\theta}}~,\label{eq:semiclassical entropy}
\end{equation}
where $S_0$ is a constant. Note that the saddle point value of $\theta$ is determined by the inverse temperature in the HH state \cite{Goel:2023svz}. \eqref{eq:semiclassical entropy} is interpreted as a semiclassical black hole entropy in sine dilaton gravity \cite{Blommaert:2024whf,Blommaert:2024ymv}, where $\Phi_h=\theta$ is the dilaton evaluated at the black hole horizon, and which is different from the BH entropy \cite{Blommaert:2024whf}.}

\paragraph{Partially-Entangled Thermal States}{The next natural generalization of the above results is to consider the one-particle generalization of the HH state \eqref{eq:HH state tL tR} $\rme^{-\frac{\beta_L}{2}\hH_L-\frac{\beta_R}{2}\hH_R}\hmO_\Delta^L\ket{\Omega}$ \eqref{eq:HH state tL tR}. Its density matrix \eqref{eq:rho from state definition} is
\begin{equation}\label{eq:density matrix PETS}
\begin{aligned}
        \hrho_L&=\rme^{-\frac{\beta_R}{2}\hH_L}\hmO_\Delta^L\rme^{-\beta_L\hH_L}\hmO_\Delta^L\rme^{-\frac{\beta_R}{2}\hH_L}\\
    &=\rme^{-{\beta_R}\hH}q^{\Delta\hat{n}}\rme^{-\beta_L\hH}~,
\end{aligned}
\end{equation}
where the second line follows from \eqref{eq:partition 1 particle}. The right-hand-side in the first line is precisely the chord version of the density matrix of a partially-entangled thermal states (PETS) (5.4) in \cite{Goel:2023svz}. The difference is that in their construction, they evaluate the entanglement entropy between a pair of SYK models with operator insertions, while here we evaluate entanglement entropy from the double-scaled algebra of a single DSSYK in its chord diagram formulation. While these are different physical systems, they are related through a Choi-Jamiołkowski \cite{Choi:1975nug,Jamiolkowski:1972pzh} isomorphism after implementing the double-scaling limit and with annealed averaging in the physical SYK pair, as we will detail in an upcoming work. 
\\
Next, we can evaluate the entanglement entropy formula \eqref{eq:algebraic entropy} from \eqref{eq:density matrix PETS}; however, it is more convenient to work with the Rényi entropy \eqref{eq:algebra Renyi entropy}, which gives
\begin{subequations}\label{eq:Renyi entropy}
    \begin{align}
     S_\alpha&\equiv \qty(1-\alpha\pdv{\alpha})\frac{\bra{\Omega}\qty(\rme^{-{\beta_R}\hH}q^{\Delta\hat{n}}\rme^{-\beta_L\hH})^\alpha\ket{\Omega}}{\qty(Z_\Delta(\beta_L,\beta_R))^\alpha}\label{eq:entropy rhoalpha}\\
     &=\qty(1-\alpha\pdv{\alpha})Z_{\Delta}(\beta_L,\beta_R)^{-\alpha}\prod_{i=0}^{\alpha-1}\int\rmd\theta_i\mu(\theta_i)\rme^{-{\beta_{i}}E(\theta_i)}\frac{\bra{\theta_i}q^{\Delta\hat{n}}\ket{\theta_{i+1}}}{\qty(Z_{\Delta}(\beta_L,\beta_R))^{\alpha}}~,\label{eq:new expression rho n}
\end{align}
\end{subequations}
where $Z_\Delta(\beta_L,\beta_R)$ appears in \eqref{eq:partition 1 particle}, and
\begin{equation}
    \beta_i=\begin{cases}
        \beta_R~,&i=2\mathbb{Z}_{\geq0}~,\\
        \beta_L~,&i=2\mathbb{Z}_{\geq0}+1~,
    \end{cases}
\end{equation}
and $\mathbb{Z}$ denotes the integers. The expression \eqref{eq:entropy rhoalpha} is precisely (5.6) in \cite{Goel:2023svz}, where \eqref{eq:new expression rho n} can be equivalently evaluated using the replica trick. The result in the IR triple-scaling limit can be interpreted as a black hole entropy in dual JT gravity with matter operator insertions, as found in the PETS interpretation of \cite{Goel:2018ubv}.}

\paragraph{Fixed Chord Number State}{Another case of interest is when $\ket{\psi}=\ket{n}=:\hat{h}^n:\ket{\Omega}$, with $\hat{h}$ defined in \eqref{eq:def h}, so that we identify $\hrho=(:\hat{h}^n:)^2$ in the double-scaled algebra, and the von Neumann entropy \eqref{eq:entropy probability dist} then reduces to
\begin{equation}\label{eq:entropy n}
    S=-\int_0^\pi\rmd\theta\mu(\theta)\frac{H_n(\cos\theta|q)^2}{(q;q)_n}\log\frac{H_n(\cos\theta|q)^2}{{(q;q)_n}}~.
\end{equation}
The above formula also appeared in \cite{Tang:2024xgg} albeit without proper justification to associate it with von Neumann entropy in the double-scaled algebra, which follows from our arguments. We will discuss these expressions in more detail in Sec.~\ref{sec:RT dS2}.\footnote{{A natural extension would be to consider the density matrix associated to the chord state with an operator insertion $\ket{\Delta;n_L,n_R}=:\hat{h}_L^{n_L}\hmO_\Delta^L\hat{h}_L^{n_R}:\ket{\Omega}$ \cite{Xu:2024hoc} where $\hat{h}_{L/R}\equiv(\sqrt{\lambda}/J)\hat{H}_{L/R}$ normal ordering takes all annihilation operators to the front, similar to \eqref{eq:normal order}. This state corresponds to a two-sided wormhole length with a worldline particle insertion in the bulk \cite{Lin:2022rbf,Aguilar-Gutierrez:2025pqp,Aguilar-Gutierrez:2025mxf}.}}
}

\section{Black hole and Cosmological Entanglement Entropy from the DSSYK}\label{sec:RT dS2}
\begin{figure}
    \centering
    \includegraphics[width=0.5\linewidth]{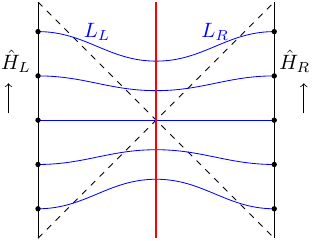}
    \caption{Effective AdS$_2$ black hole geometry in the bulk dual \cite{Blommaert:2024ymv} proposal of the DSSYK (reviewed in App.~\ref{app:sine dilaton}) {with a worldline particle, corresponding to the insertion of operators in the boundary theory.} The two-sided minimal geodesic lengths \cite{Aguilar-Gutierrez:2025sqh} (represented $L_{L/R}$, solid blue lines) evolve through boundary time, generated by the ADM Hamiltonian isomorphic to the two-sided chord Hamiltonian $\hH_{L/R}$ \eqref{eq:two_sided_DSSYK_H}. Similarly, the time-translation generator corresponds to zero-particle chord Hamiltonian \eqref{eq:H zero particle} when there are not matter operator insertions, corresponding to no particles in bulk. The dashed line represents the effective AdS$_2$ black hole horizon.}
    \label{fig:bulk_picture}
\end{figure}

In this section, we study algebraic entanglement entropy in a special limit where we zoom in the {IR and UV} limits in the energy spectrum of the DSSYK, and we match it to the area of the RT surfaces (which are spacetime points) {in JT} and dS JT gravity {respectively. The dS$_2$ limit is a new contribution, which builds on the AdS$_2$ case studied by \cite{Tang:2024xgg}, where our work justifies the interpretation of their results in terms of algebraic entanglement entropy.} Specifically, we show that the algebraic entanglement entropy for a fixed chord number in the IR and UV limits corresponds to a codimension-two area measured by the value of the dilaton at the extremal surface in JT and dS JT gravity respectively; realizing the (A)dS$_2$/CFT$_1$ correspondence explicitly. The entangling region is {at the asymptotic boundary of AdS$_2$ or} $\mathcal{I}^\pm$ in dS$_2$ respectively, as displayed in Fig.~\ref{fig:dS2_particle}. Our results are consistent with the putative duality between sine dilaton gravity and the DSSYK model (some other evidence can be found e.g.~\cite{Blommaert:2024whf,Blommaert:2024ymv,Cui:2025sgy}), which is not an input in our computations. {Particularly,} while the entangling regions are time-like separated regions in the dS$_2$ geometry, this corresponds to entanglement between space-like separated regions in the effective AdS$_2$ geometry of sine dilaton gravity, {{which}} is useful for interpreting the results in this section. {{The bulk representation of this construction is displayed in Fig.~\ref{fig:bulk_picture}.}}

We also identify that the exponent in the holographic entanglement entropy is proportional to Krylov complexity for the zero-particle HH state \eqref{eq:HH state}.

\paragraph{Outline}In Sec.~\ref{ssec:triple scaling} we discuss the triple-scaling limit in the DSSYK Hamiltonian that reproduces a generator of time/spatial translations along the asymptotic boundaries of {AdS$_2$/dS$_2$ space respectively.} In Sec.~\ref{ssec:EE bdry dS2} we explain how to carry out the boundary theory evaluation of entanglement entropy in the previous triple-scaling limit for a fixed chord number state. In Sec.~\ref{ssec:EE bulk dS2} we turn to the bulk picture by showing that the minimal area in the RT formula with respect to entangling surfaces at the asymptotic boundaries match the algebraic entanglement entropy in the UV limit; {we also recall the analogous AdS$_2$ calculation in \cite{Tang:2024xgg}.} Sec.~\ref{ssec: BH GH formulas} we specialize in recovering the {BH and} GH entropy formulas. In Sec.~\ref{ssec:higher dS/CFT} we analyze the results and compare them to dS/CFT in higher dimensions. In Sec.~\ref{ssec:Lloyd} we connect the results with Krylov complexity in the DSSYK and its rate of growth.

\subsection{IR and UV Triple-Scaling Limit of the DSSYK Hamiltonian}\label{ssec:triple scaling}
In this section, we define {IR and UV} triple-scaling limits in the DSSYK Hamiltonian that {respectively} lead to the JT and dS JT gravity generators of {time and} spatial translations along the asymptotic boundaries of {an AdS$_2$ black hole, and} $\mathcal{I}^\pm$ in dS$_2$ space (see (55) in \cite{Heller:2025ddj}). {The IR case is simply called the triple-scaling of DSSYK in the literature \cite{Lin:2022rbf}, which is used to recover the ADM Hamiltonian of JT gravity from the DSSYK chord Hamiltonian. However, here we need to distinguish two regimes associated to the minimum and maximum edges in the chord energy spectrum \eqref{eq:energy}. The latter is claimed to be} holographically dual to dS JT gravity in \cite{Blommaert:2024whf}. Note however, that the chord energy spectrum \eqref{eq:energy} is symmetric around zero energy and it is bounded (before taking a triple-scaling limit), the (A)dS JT gravity limits from the boundary perspective can be used interchangeably since one can simply rescale the Hamiltonian by an overall minus sign, {due to invariance of the chord theory under a $\mathbb{Z}_2$ transformation in the generators of the chord algebra \cite{Lin:2023trc}.}

This means that what used to the UV region (i.e.~$\theta\simeq\pi$ in \eqref{eq:energy}) becomes the IR region ($\theta\simeq0$) and viceversa by a change of sign in the Hamiltonian. Indeed, it was recently found by \cite{Heller:2025ddj} that the generator of spatial translations along $\mathcal{I}^\pm$ in dS JT gravity in appropriate canonical variables corresponds is equivalent to a opposite signed JT gravity \cite{Harlow:2018tqv}. However, while one can zoom in the sides of the energy spectrum near $\theta=0$ and $\theta=\pi$ in the same way since it is symmetric, they are still physically different limits both from the boundary and bulk perspectives. In the boundary case, these are (symmetrically) opposite sides of the spectrum; and in the bulk, they are expected to describe the JT and dS JT gravity limits of sine dilaton gravity \cite{Blommaert:2024whf}.

Based on the above context, we d{efine the IR and UV} triple-scaling limits in the DSSYK Hamiltonian as:
\begin{subequations}\label{eq:def IR UV triple scaling limit}
\begin{align}
{\rm IR~limit}\quad\rme^{-L_{\rm AdS}}&\equiv\frac{\rme^{-\lambda n}}{2\lambda^{2}}\quad\text{and}\quad {k}_{\rm AdS}\equiv\frac{\theta}{\lambda}\quad\text{are fixed as }\lambda\rightarrow0~,\\
{\rm UV~limit}\quad\rme^{\rmi L_{\rm dS}}&\equiv\frac{\rme^{-\lambda n}}{2\lambda^{2}}\quad\text{and}\quad k_{\rm dS}\equiv\frac{\pi-\theta}{\lambda}\quad\text{are fixed as }\lambda\rightarrow0~.
\end{align}
\end{subequations}
{where $L_{\rm AdS}$ and $L_{\rm dS}$} are the corresponding eigenvalues of
\begin{subequations}
\begin{align}\label{eq:dS length}
    \hat{L}_{\rm Ads}&\equiv \lambda\hat{n}+\log(2\lambda^2)\mathbb{1}~,\\
    \hat{L}_{\rm dS}&\equiv \rmi \qty(\lambda\hat{n}+\log(2\lambda^2)\mathbb{1})~.\label{eq:length dS}
\end{align}
\end{subequations} 
In the sine dilaton gravity interpretation the above lengths correspond to a regularized wormhole length in the effective AdS$_2$ geometry \cite{Blommaert:2024ymv}. We also stress that \eqref{eq:length dS} is a definition and not an analytic continuation. As mentioned in the introduction of this section, this definition of a canonical variable is used to reproduce the dS$_2$ generator of spatial translations along $\mathcal{I}^\pm$ in dS JT gravity \cite{Heller:2025ddj} from the DSSYK Hamiltonian as our definition of the UV triple-scaling limit we implement.

{Taking the IR} and UV triple-scaling limit in the zero-particle chord Hamiltonian \eqref{eq:energy basis},\footnote{See also \eqref{eq:two_sided_DSSYK_H} where we redefine the operators in terms of canonical variables \cite{Aguilar-Gutierrez:2025pqp} \begin{equation}\label{eq:non-conjugate ops many particles}
\hat{a}_{i}^\dagger=\frac{\rme^{-\rmi \hat{P}_{i}}}{\sqrt{1-q}}~,\quad \hat{\alpha}_{i}=\sqrt{1-q}~\rme^{\rmi \hat{P}_{i}}~,\quad q^{\hat{n}_{i}}=\rme^{-\hat{\ell}_{i}}~,
\end{equation}
and in the $m=0$ case $\hH_{L}=\hH_{R}\equiv\hH$ and similarly for other operators.} we get 
\begin{equation}
\label{eq:Hamiltonian simpler}
\begin{aligned}
     \hH&=-\frac{\mathcal{J}}{\lambda}\qty(\rme^{-\rmi \hP}+\rme^{\rmi \hP}\qty(1-q^{\hat{n}}))\\
    &=\begin{cases}
        \frac{\mathcal{J}}{\lambda}\qty(-2+\lambda^2\qty(\frac{\hat{{P}}_{\rm AdS}^2}{2}+2\rme^{-\hat{L}_{\rm AdS}})+\mathcal{O}(\lambda^4))~,&{\rm IR ~limit}\\
        \frac{\mathcal{J}}{\lambda}\qty(2+\lambda^2\qty(\frac{\hat{{P}}_{\rm dS}^2}{2}-2\rme^{\rmi\hat{L}_{\rm dS}})+\mathcal{O}(\lambda^4))~,&{\rm UV~ limit}
    \end{cases}
\end{aligned}
\end{equation}
where $\mathcal{J}\equiv\lambda J/\sqrt{\lambda(1-q)}$; and $\hat{P}_{\rm (A)dS}$ is the conjugate momentum to $\hat{L}_{\rm (A)dS}$. The zero-point energy subtracted Hamiltonian truncated to the first non-trivial order can be expressed
\begin{subequations}\label{eq:new Hamiltonian dS2}
\begin{align}
    {\rm IR~limit}:\quad  \frac{1}{\lambda^2}\qty(\hH+\frac{2\mathcal{J}}{\lambda}\mathbb{1})&=\frac{\mathcal{J}}{\lambda}\qty(\frac{\hat{{P}}_{\rm AdS}^2}{2}+2\rme^{-\hat{L}_{\rm AdS}}+\mathcal{O}(\lambda^2))~,\\
    {\rm UV~limit}:\quad  \frac{1}{\lambda^2}\qty(\hH-\frac{2\mathcal{J}}{\lambda}\mathbb{1})&=\frac{\mathcal{J}}{\lambda}\qty(\frac{\hat{{P}_{\rm dS}}^2}{2}-2\rme^{\rmi\hat{L}_{\rm dS}}+\mathcal{O}(\lambda^2))~
\end{align}
\end{subequations}
Truncating the above relations to the first non-trivial order and expressing them in terms of \eqref{eq:length dS} and its conjugate momentum $\hat{P}_{\rm dS}$ for the UV limit, we have
\begin{subequations}\label{eq:dS Hamiltonian}
\begin{align}
{\rm IR~limit}:\quad    \hH_{\rm AdS}&\equiv \eval{\frac{1}{\lambda^2}\qty(\hH+\frac{2\mathcal{J}}{\lambda}\mathbb{1})}_{\mathcal{O}(\lambda^{-1})}= \frac{J}{\lambda}\qty(\frac{\hat{{P}}_{\rm AdS}^2}{2}+2\rme^{-{\hat{L}_{\rm AdS}}})~,\\
{\rm UV~limit}:\quad      \hH_{\rm dS}&\equiv \eval{\frac{1}{\lambda^2}\qty(\hH-\frac{2\mathcal{J}}{\lambda}\mathbb{1})}_{\mathcal{O}(\lambda^{-1})}=\frac{J}{\lambda}\qty(\frac{\hat{{P}}_{\rm dS}^2}{2}-2\rme^{\rmi{\hat{L}_{\rm dS}}})~,
    \end{align}
\end{subequations}
where $\hH_{\rm dS}$ corresponds to the generator of spatial translations along $\mathcal{I}^\pm$ in dS$_2$ space \cite{Heller:2025ddj} (55) which we recover from a boundary evaluation; {and $\hH_{\rm AdS}$ the ADM Hamiltonian of JT gravity at the disk level \cite{Harlow:2018tqv}.} 
The DSSYK energy spectrum then becomes
\begin{equation}\label{eq:enerrgy spectrum small}
    E(\theta)=-\frac{2\mathcal{J}}{\lambda}\cos\theta=\begin{cases}
        \frac{2\mathcal{J}}{\lambda}\qty(-1+\frac{\lambda^2{k}_{\rm AdS}^2}{2}+\mathcal{O}(\lambda^4))&{\rm IR ~limit}~,\\
        \frac{2\mathcal{J}}{\lambda}\qty(1-\frac{\lambda^2{k}_{\rm dS}^2}{2}+\mathcal{O}(\lambda^4))&{\rm UV~ limit}~.
    \end{cases}
\end{equation}
Note $\eta=+1$ corresponds to the IR limit of the DSSYK energy spectrum, and $\eta=-1$ the UV; and that the DSSYK Hamiltonian \eqref{eq:dS Hamiltonian} is always Hermitian (with respect to the chord inner product \cite{Lin:2022rbf}), which is also manifest in \eqref{eq:dS Hamiltonian} since the eigenvalues of $\hat{L}$ and $\hat{L}_{\rm dS}$ are real and imaginary respectively. We stress we are not assuming a holographic correspondence in this derivation; instead, we reproduce the generator of time translations along the asymptotic boundaries in the AdS$_2$ black hole \cite{Harlow:2018tqv} and spatial translations along $\mathcal{I}^\pm$ in \cite{Heller:2025ddj}\footnote{The expression was missing $\rmi$ factors in the first preprint version.} respectively from the boundary theory Hamiltonian seen as the IR and UV limits of the DSSYK model respectively.

\subsection{Algebraic Entanglement Entropy in the UV/IR Limits}\label{ssec:EE bdry dS2}
{In this subsection, we specialize in the reduced density matrix \begin{equation}\label{eq:rho from n}
    \hrho=(:\hat{h}^n:)^2~,
    \end{equation}
which is normalized ($\Tr\hrho=\bra{n}\ket{n}=1$), the normal ordering was defined in \eqref{eq:normal order}, and $\hat{h}$ appears in \eqref{eq:def h}.} We seek to evaluate the algebraic entanglement entropy \eqref{eq:entropy probability dist} difference involving \eqref{eq:entropy n},
\begin{equation}\label{eq:important entropy difference}
\begin{aligned}
    &\Delta S\equiv -\Tr\qty(\hat{\rho}\log\hat{\rho}-\eval{\hat{\rho}\log\hat{\rho}}_{n\rightarrow\infty})~,\\
    &{\rm where~~}\Tr\qty(\hat{\rho}\log\hat{\rho})=\int_0^\pi\rmd\theta \mu(\theta) \abs{\bra{\theta}\ket{n}}^2\log(\abs{\bra{\theta}\ket{n}}^2)~,
\end{aligned}
\end{equation}
with $\bra{\theta}\ket{n}$ appears in \eqref{eq:H_n def}, and the subtraction in the last term allows $\Delta S$ to remain finite in the UV/IR triple-scaling limits. Note that while the entanglement entropy \eqref{eq:entropy probability dist} is evaluated by integrating over all $\theta$, in the IR and UV triple-scaling limits we perform a WKB approximation \cite{wentzel1926verallgemeinerung,kramers1926wellenmechanik,brillouin1926mecanique} to the integral \eqref{eq:important entropy difference} so that it captures edge of the energy spectrum {in the IR and UV} triple-scaling limit \eqref{eq:def IR UV triple scaling limit}; similar to \cite{Tang:2024xgg} in the corresponding triple-scaling. Due to symmetry under the reflection around $\theta=\pi/2$, the evaluation of \eqref{eq:important entropy difference} in both the IR and UV triple-scaling limits follows in the same way as \cite{Tang:2024xgg} (Sec 5.2). We summarize the evaluation in App.~\ref{app:WKB approx} which leads to
\begin{equation}\label{eq:k eta}
\begin{aligned}
     \Delta S\eqlambda\pi k_{\eta}~,
\end{aligned}
\end{equation}
where $k_{\eta}\geq0$ is {the value of $k_{\rm AdS}$ in the IR} or $k_{\rm dS}$ in the UV triple-scaling limit \eqref{eq:def IR UV triple scaling limit} evaluated in the WKB approximation, where there is no kinetic energy, i.e.~$\hat{P}_{\rm (A)dS}^2\simeq0$ respectively.\footnote{We stress that while we refer to the UV limit as $\theta\simeq\pi$, since $E(\theta)$ \eqref{eq:energy} reaches its maximum energy $E=2J/\lambda$; it is essentially equivalent to the IR limit near $\theta\simeq0$ where $E=-2J/\lambda$, given that they are related by an overall scaling in the Hamiltonian. This is reflected in the calculation for $\bra{n}\ket{\theta}$ \eqref{eq:H_n def} which is invariant under $\theta\rightarrow\pi-\theta$. Then in both cases the behavior of entropy difference \eqref{eq:important entropy difference} is well-approximated by the WKB approximation in \cite{Tang:2024xgg}.} The zero-point energy subtracted classical value of the conserved Hamiltonian \eqref{eq:new Hamiltonian dS2} then becomes
\begin{equation}\label{eq:new eq}
    \eval{\lambda^{-2}\qty(E+\eta \frac{2\mathcal{J}}{\lambda})}_{\mathcal{O}(\lambda^{-1})}= \eta \frac{J}{\lambda}{k}_\eta^2~,
\end{equation}
where $\eta=+1$ corresponds to the IR and $\eta=-1$ the UV triple-scaling limit. Equating \eqref{eq:new eq} with the IR and UV spectrum of the Hamiltonian \eqref{eq:dS Hamiltonian} we thus recover
\begin{equation}\label{eq:entropy Delta dS2}
    \Delta S\eqlambda\begin{cases}
    \sqrt{2}\pi\rme^{-L/2}~,&{\rm IR ~limit}\\
       \sqrt{2}\pi \rme^{\rmi L_{\rm dS}/2}~,&{\rm UV~ limit}
        \end{cases}
\end{equation}
in units where the (A)dS$_2$ cosmological constant length scale is set to unity. We stress that the trace used to evaluate the algebraic entanglement entropy above is uniquely defined up to a scaling, so the overall constant factor in \eqref{eq:entropy Delta dS2} is not relevant, while the functional dependence on $L_{\rm (A)dS}$ is relevant to match the result with a bulk calculation in the next subsection.

\begin{figure}
    \centering
    \includegraphics[width=0.35\linewidth]{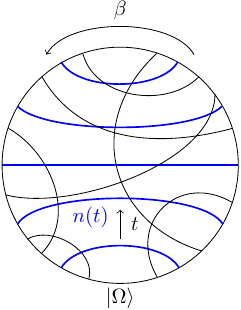}
    \caption{Representation of the semiclassical evolution of the chord number \eqref{eq:length dS} in the chord diagram (similar to \cite{Blommaert:2024ymv,Blommaert:2023opb}) prepared from the reference state $\ket{\Omega}$ and evolved in a HH preparation of state, where $n(t)$ \eqref{eq:length dS exp } counts the number of open chords for each time slice (blue solid curves), and $\beta(\theta)$ is the periodicity of the thermal circle \eqref{eq:beta semiclassical}.}
    \label{fig:thermal circle dS}
\end{figure}

\paragraph{Time Dependence of the Chord Number}
{In the above computations, we have taken $L_{\rm AdS}$ and $L_{\rm dS}$ to be fixed parameters determined from the UV and IR triple-scaling limits \eqref{eq:def IR UV triple scaling limit} of the parameter $n$ in the density matrix $\hrho=(:\hat{h}^n:)^2$ for $\hat{h}$ in \eqref{eq:def h}. We would like to study the consequences of promoting it $n$ to a time-dependent expectation value. This is naturally motivated when evaluating expectation values of operators in chord diagrams, with respect to the HH state \eqref{eq:HH state} $\rme^{-\tau\hH}\ket{\Omega}$ where $\tau=\frac{\beta}{2}+\rmi t$, as illustrated in Fig.~\ref{fig:thermal circle dS}.} 
For instance
\begin{equation}\label{eq:n HH state}
n(t)\equiv \frac{\langle\Omega|{\rm e}^{-\tau^*\hat{H}}\hat{n}{\rm e}^{-\tau\hat{H}}|\Omega\rangle}{\langle\Omega|{\rm e}^{-\beta\hat{H}}|\Omega\rangle}~,
\end{equation}
where the superscript $^*$ indicates complex conjugate,
and $\hH$ is in \eqref{eq:Hamiltonian simpler}.

In particular, in the IR and UV limits we define
\begin{equation}\label{eq:length dS exp }
\begin{aligned}
L_{\rm AdS}(t)\equiv&\frac{\langle\Omega|{\rm e}^{-\tau^*\hat{H}}\hat{L}_{\rm AdS}{\rm e}^{-\tau\hat{H}}|\Omega\rangle}{\langle\Omega|{\rm e}^{-\beta\hat{H}}|\Omega\rangle}~,\quad L_{\rm dS}(t)\equiv&\frac{\langle\Omega|{\rm e}^{-\tau^*\hat{H}}\hat{L}_{\rm dS}{\rm e}^{-\tau\hat{H}}|\Omega\rangle}{\langle\Omega|{\rm e}^{-\beta\hat{H}}|\Omega\rangle}~,
    \end{aligned}
\end{equation}
The above expectation value can be analytically computed in the semiclassical inverse temperature of the DSSYK/sine dilaton gravity \cite{Goel:2023svz,Blommaert:2024ymv} 
\begin{equation}\label{eq:beta semiclassical}
    \beta(\theta)\eqlambda\frac{\pi-2\theta}{J\sin\theta}~.
\end{equation}
The evaluation of \eqref{eq:length dS exp } then results in \cite{Heller:2024ldz,Heller:2025ddj}
\begin{equation}\label{eq:lengths dS2 time2}
L_{\rm AdS}(t)=2\log(\frac{\cosh(\theta_{\rm AdS} t)}{\theta_{\rm AdS}}) ~,\quad
L_{\rm dS}(t)=2\rmi\log(\frac{\cosh(\theta_{\rm dS} t)}{\theta_{\rm dS}}) ~,    
\end{equation}
where $\theta_{\rm dS}\equiv\pi-\theta$, equivalent to $\lambda\tilde{k}$ in the UV limit \eqref{eq:def IR UV triple scaling limit}; and {similarly for $\theta_{\rm AdS}\equiv\theta=\lambda\tilde{k}$ in the IR limit \eqref{eq:def IR UV triple scaling limit} in the above expression.}

\subsection{Minimal Area in the RT Formula for (A)dS\texorpdfstring{$_2$}{} space}\label{ssec:EE bulk dS2}
\begin{figure}
    \centering
    \includegraphics[width=0.5\linewidth]{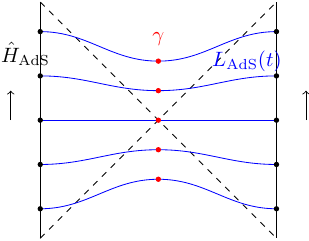}\hfill\includegraphics[width=0.37\linewidth]{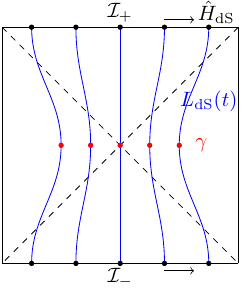}
    \caption{(A)dS$_2$ geometry where minimal length geodesic lines (blue) connect the \emph{entangling regions} (black dots) at the {timelike asymptotic boundaries (left)} or $\mathcal{I}^\pm$ (right) to the \emph{RT surfaces} (red dots, denoted $\gamma$ \eqref{eq:RT surface dS}), where the dilaton reaches is minimal value with respect to the homology constraint. The geodesics, with length $L_{\rm (A)dS}(t)$ in the diagrams, serve to gauge-invariantly define the location of RT surface {in Rindler-AdS$_2$ (left), or} in the static patch (right) as we translate (black arrows) the location of the entangling surface points (black dots) at the boundaries, which are gauge-fixed at symmetric points, though the generators {$\hH_{\rm AdS}$ (left) and} $\hH_{\rm dS}$ (right), see \eqref{eq:dS Hamiltonian}. The dashed black lines represent the black hole (left) or cosmological (right) horizons. The Milne patch in dS$_2$ space is the expanding region outside the cosmological horizon and bounded by $\mathcal{I}^\pm$; while the static patch is its complement in the global geometry.}
    \label{fig:dS2_particle}
\end{figure}

In this section, we verify that our boundary computation of entanglement entropy \eqref{eq:entropy Delta dS2} has a geometric interpretation by evaluating the codimension-two area {in JT and} dS JT gravity ({the latter one is} seen as a s-wave dimensional reduction from dS$_3$) at the extremal surface homologous to the entangling region (described below):
\begin{equation}\label{eq:RT surface dS}
    S_{\rm AdS}\equiv \frac{{\Phi_{\rm AdS}(\gamma)}}{4G_N}~,\quad S_{\rm dS}\equiv \frac{{\Phi_{\rm dS}(\gamma)}}{4G_N}~,
\end{equation}
with $G_N$ being the two-dimensional Newton's constant, $\Phi_{\rm (A)dS}$ is the dilaton in {JT and} dS JT gravity respectively (which we review in App.~\ref{app:sine dilaton}, see \eqref{eq:dS JT gravity action}),\footnote{We refer the reader to e.g.~\cite{Svesko:2022txo} for detailed definitions of dS JT gravity from s-wave reductions in higher-dimensional spacetimes. It would be interesting to study if the relation between sine dilaton and dS JT gravity can also be extended for the s-wave reduction of the near Nariai black holes \cite{Svesko:2022txo}.} and $\gamma$ is the extremal codimension-two surface, illustrated in Fig.~\ref{fig:dS2_particle}. 
Below, we confirm that the most straightforward implementation of the RT formula in {JT and} dS JT gravity matches \eqref{eq:entropy Delta dS2} and thus the RT formula corresponds to holographic entanglement entropy.

\paragraph{On the Entangling Region}The entangling region in this particular case corresponds to points at the asymptotic boundaries in (A)dS$_2$. A different possibility would have been to use finite size intervals at the boundaries as the entangling surface, which would correspond to pseudo entropy in the dS$_2$ case \cite{Doi:2022iyj}, and timelike entropy in AdS$_2$. {Both measures are equivalent at least in three-dimensions in the bulk \cite{Harper:2025lav}.} From the boundary perspective, this would correspond to using time intervals to define the entangling subregions, since it cannot have spatial dependence. For this reason, we only consider entangling region points. Nevertheless, there is still geometric information encoded in the bulk from the boundary evolution, since the RT surface points always reside inside the bulk  (Fig.~\ref{fig:bulk_picture}).

The location of the RT surface, $\gamma$, is determined from the homology constraint to the entangling surfaces, which therefore changes as the entangling surface points are translated along the asymptotic boundaries; see Fig.~\ref{fig:dS2_particle}, where we gauge-invariantly specify the location of the RT surfaces by dressing it through minimal length geodesic curves connecting the asymptotic boundaries. In the following, we gauge-fix the coordinate system to describe the location of the entangling surfaces in our evaluations. We describe the entangling surface points in {Rindler-AdS$_2$ coordinates or} in static patch coordinates for the Milne patch of dS$_2$ space (Fig.~\ref{fig:dS2_particle}) respectively as
\begin{equation}\label{eq:static patch}
\begin{aligned}
    {\rm AdS:}\quad&\rmd s^2=-(r^2-r_{\rm BH}^2)\rmd t^2+\frac{\rmd r^2}{r^2-r_{\rm BH}^2}~,\\
   {\rm dS:}\quad&\rmd s^2=-(r_{\rm CH}^2-r^2)\rmd t^2+\frac{\rmd r^2}{r_{\rm CH}^2-r^2}~,
\end{aligned}
\end{equation}
where {$r_{\rm BH}$ and $r_{\rm CH}$ are the black hole and} cosmological horizon location {respectively}. Note while for two-sided geometries we can use different $t$ coordinates, denoted $t_{L/R}$, at the asymptotic boundaries (where $r\rightarrow\infty$), we have gauge-fixed them to be the same value, i.e.~$t_{L}=t_{R}=t$. This is just a gauge-fixing choice; any choice other than $t_L=-t_R$ is gauge-equivalent by boost isometries.\footnote{This can be seen for instance from the fact that the one-particle HH state \eqref{eq:HH state tL tR} reduces to the zero-particle state \eqref{eq:HH state} when there is no operator insertion, $\hH_L=\hH_R=\hH$ in \eqref{eq:HLmultiple}. Let $t_{L/R}$ denote the coordinate $t$ along the corresponding asymptotic boundaries (i.e.~$\mathcal{I}^\pm$ in the dS$_2$ case), with generators $\hH_{L/R}$. We see that the translation operator $\rme^{-\rmi( t_L\hH_L+t_R\hH_R)}$ in the one-particle HH state \eqref{eq:HH state tL tR} reduces to $\rme^{-\rmi t\hH}$ in the zero-particle HH state \eqref{eq:HH state} by simply identifying $t=t_L+t_R$; thus, the gauge equivalence.}

\paragraph{Evaluation}We observe that {in both the AdS$_2$ and} dS$_2$ cases the RT surface points are in between the asymptotic boundaries and their location depends on the particular codimension-one slice extending between the boundaries. We can then determine its location depending on the boundary time as mentioned in the above paragraph. The details about the evaluation are straightforward from the appropriate coordinate changes in (A)dS$_2$ space; see App.~\ref{app:details evaluation RTdS2} for the dS$_2$ case and Sec.~5.3 in \cite{Tang:2024xgg} for the AdS$_2$ black hole case. The result is,\footnote{{\eqref{eq:dilaton dS terms} has the same structure as a two-point function in the semiclassical limit.}}
\begin{equation}\label{eq:dilaton dS terms}
    {\Phi_{\rm AdS}(\gamma)}=\rme^{- L_{\rm AdS}(t)/2}~,\quad{\Phi_{\rm dS}(\gamma)}=\rme^{\rmi L_{\rm dS}(t)/2}~,
\end{equation}
in units where the (A)dS$_2$ cosmological constant length scale is set to unity, and
\begin{equation}\label{eq:lengths dS2 time}
L_{\rm AdS}(t)=2\log(\frac{\cosh(r_{{\rm BH}} t)}{r_{\rm BH}}) ~, \quad L_{\rm dS}(t)=2\rmi\log(\frac{\cosh(r_{\rm CH} t)}{r_{\rm CH}}) ~,    
\end{equation}
is the regularized minimal geodesic length between points at the asymptotic boundaries, which in the  dS$_2$ case corresponds to $\mathcal{I}^\pm$ in static patch coordinates analytically continued to the Milne patch (i.e.~the expanding region in the Penrose diagram Fig.~\ref{fig:dS2_particle}, as previously explained. Comparing \eqref{eq:dilaton dS terms} with \eqref{eq:lengths dS2 time2}, we require that 
\begin{equation}\label{eq:rh thetadS}
r_{\rm BH}=\theta_{\rm AdS}~,\quad r_{\rm CH}=\theta_{\rm dS}~.    
\end{equation}
Note that generally, the RT surface points are always located in the {Rindler/}static patch of (A)dS$_2$ (Fig.~\ref{fig:dS2_particle}) when the entangling surface points are located at the {corresponding} asymptotic boundaries. In the sine dilaton gravity description, the location of the RT surface similarly corresponds to the Rindler patch of the AdS black hole entanglement entropy in the AdS$_2$ effective geometry (Fig.~\ref{fig:bulk_picture}).

Thus, we confirm there is a RT formula \cite{Ryu:2006bv,Ryu:2006ef} for (A)dS$_2$/CFT$_1$ in this concrete model. We discuss the physical interpretation and contrast {the results} with dS$_{d+1}$/CFT$_{d}$ in Sec.~\ref{ssec:higher dS/CFT}; {see also Sec.~\ref{ssec:higher(A) dS/CFT}}.

\paragraph{Relational Entropy}{In both (A)dS$_2$ cases,} the holographic entanglement entropy decreases as the entangling surface points move away from $t=0$ (i.e.~evolving through the HH state $\rme^{-\tau\hH_{\rm (A)dS}}\ket{\Omega}$), indicating that there is a maximally entangled state associated with the {BH and} GH entropies \eqref{eq:GH term}, which corresponds to $\ket{\Omega}$ ({as expected from the type II$_1$ algebra before taking the IR and UV limits \cite{Lin:2022rbf,Xu:2024hoc}}) from the physical bulk Hilbert space and boundary Hilbert space isomorphism; {at least in the absence of matter in this computation}. However, this changes for the black hole case when we increase its mass \cite{Chandrasekaran:2022cip}; while for de Sitter space the GH entropy remains as the maximal one \cite{Chandrasekaran:2022cip}. This is also consistent with the generalized horizon entropy being quantum reference frame (QRF) dependent \cite{DeVuyst:2024pop,DeVuyst:2024uvd,Araujo-Regado:2025ejs},\footnote{QRFs \cite{Krumm:2020fws,Hohn:2017cpr,Hoehn:2019fsy,Hoehn:2020epv,Hoehn:2023ehz,Vanrietvelde:2018pgb,Vanrietvelde:2018dit,delaHamette:2021oex,Hohn:2018toe,Hohn:2018iwn,Hoehn:2021flk,Giacomini:2021gei,Yang_2020,DeVuyst:2024uvd,DeVuyst:2024pop,AliAhmad:2024qrf,AliAhmad:2024vdw,AliAhmad:2024wja,DeVuyst:2025ezt,page1983evolution,Aguilar-Gutierrez:2025hty,Aguilar-Gutierrez:2025sqh,Araujo-Regado:2025ejs,Araujo-Regado:2024dpr} are field dependent frames transforming under the symmetry group of the theory; used to dress subregion operators to generate gauge invariant observables anchored to them \cite{Carrozza:2021gju,Carrozza:2022xut,Gomes:2024coh,Kabel:2023jve,Giesel:2024xtb,Araujo-Regado:2024dpr,Donnelly:2016auv}.} where we associate different spatially separated points at the asymptotic boundaries to different QRFs \cite{DeVuyst:2024pop,DeVuyst:2024uvd}. More details about the relational interpretation of the results are presented in App.~\ref{app:relational EE}. To our knowledge this is the first time where the dS space is relationally described from $\mathcal{I}^\pm$ instead of the static patch \cite{DeVuyst:2024pop,DeVuyst:2024uvd,Fewster:2024pur,AliAhmad:2024qrf}.

\subsection{Bekenstein-Hawking and Gibbons-Hawking Entropy From DSSYK}\label{ssec: BH GH formulas}
In this subsection, we analyze how to recover the BH and GH entropy formulas from previous results.

First, note that \eqref{eq:dilaton dS terms} agrees with the AdS black hole, and dS cosmological horizon respectively when the entangling surface is at $t=0$ and $r\rightarrow\infty$, thus the RT formula \eqref{eq:RT surface dS} with the explicit value of the dilaton at the minimal codimension-two surface in \eqref{eq:dilaton dS terms} reproduces the {BH entropy of a Bañados-Teitelboim-Zanelli (BTZ) black hole,} and the GH entropy \cite{Gibbons:1977mu} for dS$_3$ space \cite{Svesko:2022txo},
\begin{equation}\label{eq:GH term}
    \eval{S_{\rm AdS}}_{t=0}=\frac{2\pi r_{\rm BH}}{4 G_3}~,\quad\eval{S_{\rm dS}}_{t=0}=\frac{2\pi r_{\rm CH}}{4 G_3}~,
\end{equation}
where the three-dimensional Newton's constant, $G_3$, is related to the two-dimensional one by $G_3=2\pi G_N$ \cite{Svesko:2022txo} in units where the (A)dS length scale in the cosmological constant is set to one; also note $r_{\rm BH}$ and $r_{\rm CH}$ in the two-dimensional metric \eqref{eq:static patch} then correspond to the black hole and cosmological horizon radius of BTZ and dS$_3$ space respectively.

We also obtain the same result from the {algebraic entanglement entropy in the IR} and UV limits of the DSSYK model using (\ref{eq:entropy Delta dS2}, \ref{eq:lengths dS2 time2}) at $t=0$ since, as previously emphasized, any notion of trace is defined up to a overall rescaling by a non-zero constant. Particularly by rescaling the algebraic trace (\ref{eq:lengths dS2 time2}) with $\Tr\rightarrow \frac{1}{\sqrt{2}\pi\lambda}\Tr$ ({alternatively, by changing the} normalization of $\ket{\Omega}$) we have:
\begin{equation}\label{eq:Delta S}
    \Delta S\eqlambda\begin{cases}
    \frac{\theta_{\rm AdS}}{\lambda}~,&{\rm IR}\\
       \frac{\theta_{\rm dS}}{\lambda}~,&{\rm UV}
        \end{cases}
\end{equation}
where $\lambda=8\pi G_N$ \cite{Blommaert:2025eps},\footnote{\label{fnt:conventions}See e.g.~(2.32) \cite{Blommaert:2025eps} where $\pi {\rm \textbf{b}}^2=4G_N$ in our notation.} and $\theta_{\rm AdS}=r_{\rm BH}$, $\theta_{\rm dS}=r_{\rm CH}$ \eqref{eq:rh thetadS}. Thus, we reproduce the three-dimensional {BH and} GH entropy \eqref{eq:GH term} from the regularized entanglement entropy in the DSSYK model in the {IR and} UV triple-scaling limit. We discuss the physical significance of the results at the end of the subsection.

The above result can also be obtained from the von Neumann entropy of $\hrho=\rme^{-\frac{\beta}{2}\hH}$ in \eqref{eq:entropy} and taking {the IR/}UV triple-scaling limits, namely from
\begin{equation}\label{eq:thermal AdS dS entropy}
    S_{\rm (A)dS}^{\rm thermal}\equiv\eval{(1-\beta\partial_\beta)Z(\beta)}_{\beta=\beta_{\rm(A)dS}}~,\quad Z(\beta)=\int_{-\infty}^{\infty}\rmd s~\rho(s)\rme^{-\beta \frac{Js^2}{4\lambda}}~,
\end{equation}
where we apply the rescaling of the integration measure $\mu(\theta)\rightarrow\rho(s)$ in \eqref{eq:triple scaling}. Note that this is a boundary calculation; the bulk calculation is well-known (see review \cite{Mertens:2022irh}) and it is one-loop exact \cite{Stanford:2017thb}; and we denote $\beta_{\rm (A)dS}$ the saddle-point value $\beta(\theta)$ \eqref{eq:beta semiclassical} {in the IR ($0<\theta_{\rm AdS}\equiv\theta\ll1$)} and UV ($0<\theta_{\rm dS}(\equiv\pi-\theta)\ll1$) limits respectively; i.e.~
\begin{equation}\label{eq:semiclassical (A)dS inv temp}
    \beta_{\rm AdS}=\frac{\pi}{J\theta_{\rm AdS}}~,\quad \beta_{\rm dS}=\frac{\pi}{J\theta_{\rm dS}}~.
\end{equation}
The evaluation of the partition function \eqref{eq:thermal AdS dS entropy} is known, i.e.~\cite{Stanford:2017thb,Jafferis:2019wkd,Mertens:2022irh}, and we again recover \eqref{eq:Delta S} from \eqref{eq:thermal AdS dS entropy}, namely
\begin{equation}\label{eq:imp eq}
       S_{\rm AdS}^{\rm thermal}=\eval{\Delta S}_{t=0,{\rm IR}}~,\quad S_{\rm dS}^{\rm thermal}=\eval{\Delta S}_{t=0,{\rm UV}}~.
\end{equation}
As expected, we recover the {BH and} GH entropies from entanglement entropy in the double-scaled algebra. This is possibly not surprising since the chord Hamiltonian reduces to the JT and dS JT gravity generators \eqref{eq:dS Hamiltonian} in the IR/UV limits. We showed that there are at least two ways to do so; the arguably most intuitive one from the bulk or boundary theory {is using the density matrix $\hrho=\rme^{-\frac{\beta}{2}\hH}$ in the double-scaled algebra; and an alternative option is through $\hrho=\qty(:\hat{h}^n:)^2$ \eqref{eq:rho from n}, with $n$ taken as the expectation value of the chord number in the HH state \eqref{eq:n HH state}, since it generates RT surfaces in the bulk which intercept the black hole horizon at a particular Cauchy surface, as displayed in Fig.~\ref{fig:dS2_particle}. The connection between these approaches is thus the HH state.}\footnote{{It would be interesting to show whether the result \eqref{eq:entropy Delta dS2} beyond the minimal value for the rescaled chord number can also be interpreted from the thermal entropy formula \eqref{eq:thermal AdS dS entropy}.}}

In contrast to \eqref{eq:Delta S} and \eqref{eq:imp eq}, the GH entropy formula \cite{Gibbons:1977mu} in the higher-dimensional dS context is just a proposal to define a thermodynamics entropy associated to the cosmological horizon, and it might not have a statistical interpretation nor a microscopic description, which might also provide useful information about the dS space cosmological constant \cite{Galante:2023uyf,Harris:2023cow,Anninos:2012qw}. Regardless, \eqref{eq:Delta S} shows that at least from the lower dimensional perspective the GH entropy formula is indeed the von Neumann entropy of a thermal density matrix in the algebra of observables of the boundary dual theory;\footnote{However, to understand the microscopic origin of this formula, one would ideally like to recover an approximation to this result in the finite N analogue of the chord Hamiltonian.} which we can recover from the partition function of the UV limit of the DSSYK, or sine dilaton gravity \cite{Blommaert:2024whf}.

\subsection{Comparison with \texorpdfstring{dS$_{d+1}$/CFT$_d$}{}}\label{ssec:higher dS/CFT}
In this subsection and we compare the previous dS$_2$/CFT$_1$ results with dS/ CFT in higher dimensions; specially the RT formulas in dS$_{d+1\geq3}$ space with entangling surfaces at $\mathcal{I}^\pm$ \cite{Dong:2018cuv}.

\paragraph{Higher-Dimensional dS Holography}
From the dS$_2$ description, the result \eqref{eq:dilaton dS terms} is strikingly different from dS/CFT in higher dimensions where applying the RT formula \cite{Narayan:2022afv} with entangling surfaces between $\mathcal{I}^\pm$ in dS$_{d+1\geq3}$ results in divergent and complex-valued minimal areas due to the time-like extend of the codimension-two extremal surfaces (anchored between the two entangling surfaces) \cite{Nanda:2025tid,Narayan:2015vda,Sato:2015tta,Narayan:2024fcp,Narayan:2022afv,Goswami:2021ksw,Fernandes:2019ige,Goswami:2024vfl,Narayan:2023zen,Narayan:2023ebn,Doi:2022iyj,Narayan:2020nsc,Narayan:2019pjl,Narayan:2017xca,Narayan:2015oka}; since points at $\mathcal{I}^\pm$ cannot be connected by space or null-like geodesics. In three and higher spacetime dimensions, this signals non-unitarity in the dual CFT (due to the presence of complex conformal weights \cite{Doi:2022iyj}), assuming a form of holographic entanglement entropy in dS/CFT \cite{Strominger:2001pn}.\footnote{\label{fnt:examples dSCFT}Examples include higher spin gravity \cite{Vasiliev:1999ba,Vasiliev:1990en} where dS$_4$ space \cite{Anninos:2011ui,Ng:2012xp,Anninos:2012ft} is dual to Sp($N$) CFT$_3$s; as well dS$_3$ space being dual to a SU($2$) Wess-Zumino-Witten model \cite{Hikida:2021ese,Hikida:2022ltr,Chen:2022ozy,Chen:2022xse}.} The physical interpretation of entanglement entropy for time-like separated regions (such as in terms of pseudo entropy \cite{Doi:2022iyj} or time entanglement \cite{Narayan:2023ebn}) remains opaque to this date, although there is a lot of progress in this direction \cite{Nanda:2025tid,Narayan:2015vda,Sato:2015tta,Narayan:2024fcp,Narayan:2022afv,Goswami:2021ksw,Fernandes:2019ige,Goswami:2024vfl,Narayan:2023zen,Narayan:2023ebn,Doi:2022iyj,Doi:2022iyj,Narayan:2020nsc,Narayan:2019pjl,Narayan:2017xca,Narayan:2015oka}.\footnote{If one were to work with regularized time entanglement \cite{Narayan:2022afv} in these proposals, depending on the details, the real part can be proportional to the GH entropy which are reminiscent of the two-dimensional result in \eqref{eq:dilaton dS terms}.} For this reason, we studied a top-down example (in dS$_2$/CFT$_1$), where one can deduce the physical interpretation of entanglement between time-like separated regions. In the two-dimensional case the codimension-two extremal surfaces are points and the area, measured by the dilaton, is real; which is consistent with the entanglement entropy from the boundary perspective (i.e.~the IR and UV limits of the DSSYK model) being real-valued. In contrast with higher-dimensional proposals, the boundary theory is \emph{manifestly unitary}.

While our results above do not display some of puzzling features encountered in higher-dimensional proposals of dS/CFT mentioned above, this should not be taken as an indication that the same should hold in higher dimensions. Since the boundary theory is non-unitary (see footnote \ref{fnt:examples dSCFT}). One should properly carry out a first principles derivation of the entanglement entropy from a top-down example of dS$_{d+1\geq3}$/CFT$_{d\geq2}$, which is outside the scope of this work.

However, if instead, the boundary theory is unitary, as in this study; then the most natural extension of the results is to evaluate the RT formula in dS/CFT \cite{Lewkowycz:2019xse,Dong:2018cuv,Sato:2015tta} times an additional factor $\rmi$ when the entangling surfaces are points, which would lead to real entropies, consistent with unitarity in the dual boundary theory. This situation changes substantially when entangling surface has spatial extend (as in \cite{Doi:2022iyj,Narayan:2022afv}) since the result from the RT formula generally is not purely imaginary.

A different approach in dS holography literature is to include auxiliary time-like boundaries for dS space\footnote{\label{fnt:thermowelposed}However, when the dS stretched horizon has Dirichlet boundary conditions this leads to thermodynamic instabilities \cite{Svesko:2022txo,Banihashemi:2022jys,Banihashemi:2022htw,Anninos:2024wpy} (and it is not well-posed in general relativity \cite{Anninos:2024wpy}). Meanwhile, the system is thermodynamically stable (and well-posed) for conformal boundary conditions \cite{Anninos:2024wpy,Galante:2025tnt}; which leads to difficulties in defining dynamical dressed observables with respect to the asymptotic boundary since it has dynamical gravity, and (like other boundary conditions) they are not universally well-posed \cite{Liu:2025xij,Liu:2024ymn}.} (which we refer to as a cosmological stretched horizon \cite{Susskind:2021esx}) allowing for a more direct analogy with the RT formula in AdS/CFT to dS space \cite{Ruan:2025uhl,Franken:2023jas,Murdia:2022giv,Nomura:2017fyh,Arias:2019zug,Arias:2019pzy,Franken:2023pni,Franken:2024ruw}. However, this approach leads to some inconsistencies (pointed out in \cite{Chang:2024voo,Kawamoto:2023nki}) unless one modifies the RT formula itself in a way that involves additional free parameters (which can be bounded by requiring certain entropic inequalities are satisfied \cite{Ruan:2025uhl}). Since the dS spacetime with Dirichlet time-like boundaries is thermodynamically unstable and not well-posed \cite{Anninos:2024wpy}, we have focused on the $\mathcal{I}^\pm$ perspective (i.e.~without auxiliary boundaries); while there are possible connections between these approaches (discussed in Sec.~\ref{ssec:outlook bulk}).

\paragraph{Thermal Ensembles}
To make more contrast with higher dimensions, note that the thermal ensemble in boundary theory is well-defined even in UV limit of the DSSYK partition function \eqref{eq:partition 1 particle}. While this is natural from the AdS$_2$ black hole being prepared in a HH state, it is a priori difficult to interpret in the dS$_2$ bulk since it lacks auxiliary time-like boundaries to generate the HH preparation of state. One might need to rely on a path integral proposal to be able to interpret the boundary ensemble, as it is done in higher-dimensional dS space \cite{Gibbons:1977mu,Gibbons:1976ue}, which a priori has no statistical description. However, the bulk thermal ensemble is still well-defined in the effective AdS$_2$ geometry of sine dilaton gravity. The natural microcanonical temperature in the bulk is the fake temperature \cite{Lin:2023trc} instead, which encodes the microcanonical temperature of the boundary theory \cite{Blommaert:2024ymv}.

\subsection{Spread Complexity \& Entanglement Entropy}\label{ssec:Lloyd}
Returning to the (A)dS$_2$ interpretation (displayed in Fig.~\ref{fig:dS2_particle}) of the algebraic entanglement entropy in the UV {and IR} limits \eqref{eq:entropy Delta dS2}, we note that the entropy formula has a simple dependence in terms of minimal geodesic lengths \eqref{eq:lengths dS2 time}  (i.e.~the rescaled expectation value of the chord number) at least in the semiclassical limit, given in \eqref{eq:dilaton dS terms}. From the boundary side, we selected the expectation value of the chord number operator to be evolve in the HH state semiclassically, which corresponds to Krylov spread complexity of the zero-particle HH-state \cite{Rabinovici:2023yex,Heller:2024ldz} in {the IR/}UV triple-scaling limits \eqref{eq:def IR UV triple scaling limit} to match to the minimal area computed through the RT formula, where Krylov complexity is equivalent to a length, as discussed in several works (e.g.~\cite{Heller:2024ldz,Heller:2025ddj,Rabinovici:2023yex,Ambrosini:2025hvo,Aguilar-Gutierrez:2025mxf,Aguilar-Gutierrez:2025hty,Aguilar-Gutierrez:2025pqp,page1983evolution,Fu:2025kkh}). Thus, as seen from \eqref{eq:entropy Delta dS2} the entanglement entropy is directly related to spread complexity in this specific case where the algebraic entanglement entropy of the DSSYK model has a bulk interpretation.

We also note that the exponent of the dS entanglement entropy is determined by 
\begin{equation}\label{eq:speed bound dS HEE}
    \frac{1}{\Delta S}\dv{\Delta S}{t}=\begin{cases}
    -\frac{1}{2}\dv{L}{t}~,&{\rm IR}~,\\
    \frac{\rmi}{2}\dv{L_{\rm dS}}{t}~,&{\rm UV}~.
    \end{cases}
\end{equation}
The late time rate of growth of Krylov spread complexity therefore determines a lower bound on the growth of entanglement entropy in the IR/UV limits. We analyze this perspective in App.~\ref{app:details speed limits} using an analogous Lloyd bound for Krylov spread complexity at the semiclassical level and including first order quantum corrections using the Robertson uncertainty relation which determines speed limits \cite{Hornedal:2022pkc} in the Krylov spread complexity.

\section{Discussion}\label{sec:disc}
\paragraph{Summary}
In brief, {we investigated the entanglement entropy between the double-scaled algebras in the DSSYK model given a generic chord state.}

By specializing to the HH state, we noticed that the entanglement entropy reduces to a statistical entropy. {The expressions encode information about a two-sided black hole dual, which we associated to sine dilaton gravity.} We confirmed (instead of assuming) the relation between the DSSYK model and sine dilaton gravity when we focus on a triple-scaling limit that reproduces (A)dS$_2$ space in the bulk. 
We found that the (A)dS/CFT incantation of the RT formula (involving time-like geodesics in the bulk to evaluate the dilaton) exactly reproduces the entanglement entropy in the corresponding triple-scaling limit of the DSSYK model. In terms of QRFs, there are different holographic entanglement entropies depending on the spatial location of the entangling surface at the asymptotic boundaries. A specific entangling surface at $\mathcal{I}^\pm$ recovers the GH entropy; while different ones lead to a smaller entanglement entropy, which consistent with the type II$_1$ algebra of the DSSYK \cite{Xu:2024hoc,Lin:2022rbf,Tang:2024xgg} and with the relational interpretation of entanglement entropy \cite{Araujo-Regado:2025ejs,DeVuyst:2024pop,DeVuyst:2024uvd} (even though we are describing it relative to $\mathcal{I}^\pm$ in contrast to the previous literature, focused on the static patch \cite{DeVuyst:2024pop,DeVuyst:2024uvd,Fewster:2024pur,AliAhmad:2024qrf}).

We stress that the boundary theory remains unitary despite being placed at the $\mathcal{I}^{\pm}$ boundaries of dS$_2$ space. We see that the effective AdS$_2$ geometry takes a central role for interpreting the results. For instance, the DSSYK Hamiltonian in the triple-scaling limit \eqref{eq:dS Hamiltonian} corresponds to the time translation generator in the effective AdS$_2$ geometry. This also allows a bulk interpretation of the entanglement entropy as being computed with respect to space-like separated asymptotic boundaries in the effective geometry (which are interpreted as time-like separated in dS$_2$), and thus it obeys the basic properties as holographic entanglement entropy in AdS/CFT \cite{Nishioka:2009un,Rangamani:2016dms,Chen:2019lcd}. We also emphasize the effective geometry is central to properly define thermal ensembles without introducing auxiliary timelike surfaces in the static patch \cite{Banihashemi:2022jys}, which is one of the reasons that the gravitational path integral was proposed to define the partition function \cite{Gibbons:1976ue,Gibbons:1977mu}. However, this type of proposal is not justified to define a thermal ensemble in the sense of statistical mechanics \cite{Banihashemi:2024weu}. Our approach shares new lights in this front. The thermal ensemble in the boundary theory is mapped to the effective AdS$_2$ black hole, where the physical DSSYK temperature is replaced by that in the fake disk, and the ensemble remains well-defined.

We now comment on future research directions{, and particularly, higher dimensional extensions below}.

\subsection{Lessons for Higher-Dimensional (A)dS/CFT}\label{ssec:higher(A) dS/CFT}
{In this subsection, we are interested formalizing some of the previous findings on entanglement entropy for higher-dimensional (A)dS/CFT. In the bulk, we assume there is a cosmological constant
\begin{equation}
\Lambda=\eta\frac{d(d-1)}{2\ell^2}
\end{equation}
where $\eta=+1$ in AdS; $\eta=-1$ in dS, and $\ell$ is a constant. $\Lambda$ appears in the Einstein-Hilbert action as
\begin{equation}
I=-\frac{1}{16\pi G_N}   \int\rmd^dx\sqrt{\abs{g}}\qty(\mathcal{R}-2\Lambda)+I_{\rm bdry}~.
\end{equation}
We shall study asymptotically (A)dS$_{d+1}$ spacetimes,\footnote{We have presented a related construction in \cite{Aguilar-Gutierrez:2023hls} for axion-de Sitter universes assuming a form of the dS/CFT correspondence \cite{Strominger:2001pn}; which we expand and detail more in this work, including its AdS/CFT counterpart.} where the induced metric at the asymptotic boundaries $\Sigma_{A/B}$ (corresponding to $\mathcal{I}^\pm$ in the dS case) can be put in the form
\begin{equation}
    \eval{\rmd s^2}_{\Sigma_{A/B}}\equiv\gamma_{ab}\rmd x^a\rmd x^b\simeq-\qty(\eta\qty(\frac{r_0}{\ell})^2+\mathcal{O}\qty(\qty(\frac{r_0}{\ell})^0))\rmd t_{A/B}^2+r_0^2\rmd\Omega_{d-1}^2~,
\end{equation}
where we consider $r_0/\ell\gg1$, with $r_0$ denoting a radial regulator of the asymptotic boundary location. Note that $t_{A/B}$ are space-like coordinates in the dS case; time-like in AdS.
We denote ${H}^{(A/B)}_{\rm bulk}$ as the generator of the associated $t_{A/B}$ translations in the bulk; explicitly
\begin{equation}\label{eq:generator bulk}
    H^{(A/B)}_{\rm bulk}=\int_{\Sigma_{A/B}}\rmd^d x\sqrt{-\gamma}~T^{t}_{t}~,
\end{equation}
where $T^t_t$ is the $t$-$t$ component (where we suppressed the ${A/B}$ subindex) of the boundary stress tensor 
\begin{equation}
    T_{ab}\equiv\frac{2}{\sqrt{\abs{\gamma}}}\frac{\delta I_{\rm bdry}}{\delta\gamma^{ab}}~.
\end{equation}
Given that the holographic correspondence is best established with large but finite number of degrees of freedom \cite{Maldacena:1997re}; in contrast to the previous sections, we will consider the boundary theory dual of asymptotically (A)dS spacetimes can be described exclusively by states with discrete spectrum, which correspond to wormholes and higher genus topologies in the gravitational path integral. These include the energy spectrum in the AdS case/ the spectrum of the translation generator at $\mathcal{I}^\pm$ in the dS case; and other quantum number corresponding to a complete set of commuting observables in general. We are particularly interested in the entanglement entropy between the CFT duals at each asymptotic boundary assuming factorizable Hilbert spaces, instead of the algebraic entanglement entropy in the previous sections.
\\
By performing canonical quantization, $H_{\rm bulk}^{A/B}$ above can be promoted to an operator in the quantum gravity theory, which might suffer from renormalization problems and in general requires a UV completion. Assuming there is a (A)dS/CFT correspondence, there exists a dual operator in the boundary theory $\hH^{(A/B)}_{\rm bdry}$, which corresponds to the CFT Hamiltonian in the asymptotically AdS case; and a generator of spatial translations in dS space.
\\
From now on, we gauge-fix $t_A=t_B$ and omit the label $A/B$. Let $\ket{E_Q^{(L/R)}}$ represent left-right eigenstates of $\hH_{\rm bdry}$, defined as
\begin{equation}
     \hH_{\rm bdry}=\sum_Q E_Q\ket{E^{(L)}_Q}\bra{E^{(R)}_Q}~,
\end{equation}
where $Q$ represents a set of quantum numbers corresponding to a complete set of commuting observables, we pick a normalized basis $\bra{E^{(R)}_Q}\ket{E^{(L)}_{Q'}}=\delta_{QQ'}$ so that
\begin{equation}
\begin{aligned}
    \hH_{\rm bdry}\ket{E^{(L)}_Q}=E_Q\ket{E^{(L)}_Q}~,\quad \bra{E^{(R)}_Q}\hH_{\rm bdry}=E_Q\bra{E^{(R)}_Q}~,
\end{aligned}
\end{equation}
and we need to distinguish between left and right when $\hH_{\rm bdry}$ is non-Hermitian, corresponding to the $\eta=-1$ case, as seen from the dual bulk generator \eqref{eq:generator bulk}).
\\
From postulates in quantum mechanics, we assume that $\ket{E_Q^{(L/R)}}$ is a complete basis since it corresponds to an energy basis the $\eta=+1$ case, and our results in the previous sections suggest that the operator $\hH_{\rm bdry}$ is also one of the complete commuting set of observables relevant to describe boundary theory in dS/CFT. We then define the Hilbert space of the boundary theories on $\Sigma_{A/B}$ as
\begin{equation}
    \mathcal{H}_{A/B} \equiv{\rm span}\qty{\ket{E^{(L)}_Q}_{A/B},~\ket{E^{(R)}_Q}_{A/B}}~.
\end{equation}
Thus, the Hilbert space associated to the boundary theory duals in two-sided asymptotically (A)dS spacetimes takes the form
\begin{equation}\label{eq:full bdry theory}
    \mathcal{H}_{\rm bdry~full}=\mathcal{H}^{(Q)}_{A}\otimes\mathcal{H}^{(Q)}_{B}~.
\end{equation}
There are entangled states from $\mathcal{H}_{A/B}$ such as
\begin{equation}\label{eq:EE}
\begin{aligned}
    |\Psi^{(L)})\equiv \sum_{Q,Q'}f_1(E_Q,\,E_{Q'})\ket{E_Q^{(L)}}_A\otimes\ket{E^{(L)}_{Q'}}_B~,\\
    (\Psi^{(R)}|\equiv \sum_{Q,Q'}f_2(E_Q,\,E_{Q'})\bra{E_Q^{(R)}}_A\otimes\bra{E^{(R)}_{Q'}}_B~,
\end{aligned}   
\end{equation}
where we emphasize $\ket{E^{(L/R)}_Q}$ is independent of $\ket{E_{Q'}^{(L/R)}}$, unless we impose an equal energy constraint in the full boundary Hilbert space \eqref{eq:full bdry theory}.
\\
From here, we can trace out one of the boundary theories to evaluate the corresponding density matrix as,\footnote{This reduced density matrix is similar to those used in pseudo entropy \cite{Doi:2022iyj,Harper:2025lav}}
\begin{equation}\label{eq:total F}
\begin{aligned}
    \hat{\rho}_{A}=\tr_B(|\Psi^{(L)})(\Psi^{(R)}|):&=\sum_{Q}\qty(\bra{E_Q}\otimes\mathbb{1})|\Psi^{(L)})(\Psi^{(R)}|\qty(\ket{E_Q}\otimes\mathbb{1})\\
    &=\sum_{Q_1,Q_2,Q_3}f_1(E_{Q_1},E_Q)f_2(E_{Q_2},E_Q)\ket{E^{(L)}_{Q_1}}_A\bra{E^{(R)}_{Q_2}}_A~.
\end{aligned}
\end{equation}
While the discussion is general at this point; to evaluate entanglement entropy of a pure bipartite state, we simply take $(\Psi^{(R)}|=(|\Psi^{(L)}))^\dagger$; while to describe dS space, they are a priori independent since $\hH_{\rm bdry}$ is non-Hermitian for $d\geq2$ as seen in the dual generator \eqref{eq:generator bulk}.
\\
Finally, we can evaluate entanglement entropy with,\footnote{Note however, if the density matrix is non-Hermitian, the overall normalization of the density matrix is not physically meaningful; the aim of this formulation is to generalize our results based on the two-dimensional results in the previous section, which we discuss further below.}
\begin{equation}\label{eq:reduced F}
    S=\log\tr_A(\hat{\rho}_A)-\frac{\tr_A(\hat{\rho}_A\log\hat{\rho}_A)}{\tr_A(\hat{\rho}_A)}~,
\end{equation}
where note we are implementing a Hilbert space trace; or equivalently, a type I algebra trace, in contrast to Secs.~\ref{sec:Algebra Entropy} and \ref{sec:RT dS2} where the corresponding subsystems A/B could not be described in terms of type I von Neumann algebras 
\\
The aim of the past arguments is to describe what is Hilbert space structure in the boundary theories for (A)dS holography in a unified perspective. Holographically, we expect that \eqref{eq:reduced F} corresponds to the extremal area given by the quantum-corrected \cite{Faulkner:2013ana,Engelhardt:2014gca,Banerjee:2010qc,Banerjee:2011jp,Mandal:2010cj,Sen:2012cj,Sen:2012dw,Sen:2012kpz,Bhattacharyya:2012wz,H:2023qko} RT formula
\begin{equation}
    S_{\rm bulk}=\frac{{\rm Area}(\gamma)}{4G_N}+\mathcal{O}(G_N^0)~,
\end{equation}
where $\gamma$ is the minimal codimension-two extremal surface homologous to the entangling region in the asymptotic boundaries $\Sigma_{A/B}$; and we included $\mathcal{O}(G_N^0)$ to represent corrections to the area formula due to finite $N$ effects in the boundary theory corresponding to finite $G_N$ corrections in the bulk, and we allow matter contributions in the holographic entanglement entropy. To confirm our arguments in the asymptotically dS space, one should apply our construction of density matrices in explicit examples of dS$_{d+1\geq3}$/CFT$_{d\geq2}$ confirm bulk Hilbert space factorization; and evaluate the entanglement entropy as described above. We hope to address this more explicitly in the future.}

\subsection{Outlook: Boundary Perspective}\label{ssec:outlook bdry}
\paragraph{Quantum Groups}
{We studied a notion of entanglement entropy associated to the double-scaled algebra and its commutant. One of the special cases of the general entanglement entropy formula \eqref{eq:algebraic entropy}, namely when there are no matter operators \eqref{eq:entropy probability dist} is very reminiscent to a different one associated to the quantum group of the DSSYK model \cite{Belaey:2025ijg} (5.1). It would be interesting to relate our works since it has been noticed that there is a connection between the double-scaled algebra and the quantum group of the DSSYK model \cite{vanderHeijden:2025zkr}.}

\paragraph{Symmetry Resolved Entanglement Entropy}
{Another interesting extension of the results is to define the double-scaled algebra for the $\mathcal{N}=2$ \cite{Berkooz:2020xne,Boruch:2023bte,Aguilar-Gutierrez:2025sqh,Watanabe:2025rwp} and other U$(1)$ charged DSSYK models \cite{Berkooz:2020uly,Narayan:2023wlk,Forste:2025gng,Gubankova:2025gbx}. One should identify the cyclic separating state of the single $\mathcal{N}=2$ and the U$(1)$-charged DSSYKs. These evaluations could be useful to extend our study in the presence of conserved charges, which can be handled using different quantum correlation measures such as symmetry-resolved entanglement entropy \cite{Monkman:2023hup}.}

\paragraph{Other Entanglement Measures}To study multipartite entanglement measures in the DSSYK model, one may need to consider different particle flavors since there is no spatial extend in the real time evolving system. The notion of color entanglement in \cite{Iizuka:2024die} has interesting characteristics, including a series of inequalities that are very similar to those for multipartite holographic entanglement entropy in higher dimensions. In the DSSYK case, we found that matter chords allow us to bypass conceptual problems for defining entanglement entropy based on the double-scaled algebra, without having to introduce chords of different flavors. It would be worth to investigate color entropy in the DSSYK model based on previous works in the analogous finite $N$ SYK \cite{Iizuka:2024die}, since there are different double-scaled models in the literature with different flavors \cite{Berkooz:2024ifu,Gao:2024lem,Berkooz:2024evs,Berkooz:2024ofm}, which display phase transitions. In particular, it would be interesting to investigate the series of inequalities developed by \cite{Iizuka:2024die} in DSSYK-type of models with more flavors, and study if there is a geometric interpretation for the inequalities in the corresponding bulk dual theory.

Another approach is to consider entropy differences between states in a time-band \cite{Jensen:2024dnl}, which may be associated with entanglement in spatial subregions in \cite{Mertens:2025rpa}. It would be interesting to extend our study of entanglement entropy between algebras based on the framework in \cite{Jensen:2024dnl}. 

\subsection{Outlook: Bulk Perspective}\label{ssec:outlook bulk}
\paragraph{Edge Modes in the Bulk}
{As mentioned in Sec.~\ref{sec:intro} in other examples in gauge theories and gravity, one constructs a factorization map using edge modes attached to the entangling surface. Edge modes are additional degrees of freedom labeled by quantum numbers and superselection sectors in the corresponding representation of the gauge group. They result from mapping between the physical Hilbert space and an extended Hilbert space (corresponding to the different superselection sectors) \cite{Jafferis:2019wkd}. The contribution of edge modes to the entanglement entropy in gravity has been studied in different contexts, see e.g.~\cite{Casini:2013rba,Donnelly:2011hn,Donnelly:2014gva,Donnelly:2014fua,Donnelly:2015hxa, Donnelly:2016auv,Wong:2017pdm,Frenkel:2023yuw,Ball:2024hqe,Fliss:2025kzi, Lin:2018xkj,Lin:2017uzr,Blommaert:2018iqz,Lin:2021tlr,Jafferis:2019wkd,Mertens:2022ujr} for an incomplete list. Given that the edge modes are associated with gauge-fixing in the kinematical bulk Hilbert space of the bulk theory \cite{Araujo-Regado:2025ejs}, they are not present in our boundary construction of the factorization map. One could study how our results relate to the literature on edge modes, for instance by investigating the operator algebra of the bulk theory before gravitationally dressing matter operators to the asymptotic boundary. After the gravitational dressing, the algebra of operators in the DSSYK model and sine dilaton gravity should be treated in a similar footing since they act on isomorphic Hilbert spaces. Thus, our results could also be interpreted in terms as gravitational entropy in sine dilaton gravity from its algebra of observables. We hope to develop this in the future.}

\paragraph{Static Patch Holography}
While our work discusses a dS/CFT description of holographic entanglement entropy (of the same form as in higher dimensions \cite{Lewkowycz:2019xse,Dong:2018cuv,Sato:2015tta}), there is also a lot of interest in defining entanglement entropy (and holographic complexity \cite{Jorstad:2022mls,Baiguera:2023tpt,Anegawa:2023wrk,Anegawa:2023dad,Auzzi:2023qbm,Aguilar-Gutierrez:2024rka,Baiguera:2024xju,Aguilar-Gutierrez:2023zqm,Aguilar-Gutierrez:2023pnn,Aguilar-Gutierrez:2024rka,Aguilar-Gutierrez:2023tic}) from a static patch perspective \cite{Susskind:2021esx,Shaghoulian:2022fop,Franken:2025lfy,Franken:2024ruw,Franken:2023pni,Anninos:2011af,Tietto:2025oxn}. 

One of the main results in this work is the bulk interpretation of entanglement entropy of the DSSYK model in the dS$_2$ limit.
It would be interesting to develop the interpretation of our results in the dS$_3$ space interpretation of the DSSYK and complex Liouville string \cite{Collier:2024kmo,Collier:2024kwt,Collier:2024lys,Collier:2024mlg,Collier:2025lux,Collier:2025pbm}/sine dilaton gravity based on the evidence in \cite{Blommaert:2025eps}. Particularly to translate entanglement entropy in the DSSYK with matter to the dS$_3$ geometry. Our results show that for a specific entangling region, the entanglement entropy in dS$_2$/CFT$_1$ encodes the GH entropy in dS$_3$ static patch holography. More generally,  we expect that the minimal length geodesic lengths in this work correspond to other geodesics connecting the antipodal static patch observers in dS$_3$ space on findings in \cite{Verlinde:2024znh} where the chord number in the boundary theory translates into static patch time differences. This would imply that while the entanglement entropy of the DSSYK is closely related to dS$_2$/CFT$_1$ in the sine dilaton gravity description; the one in the dS$_3$ bulk might be more closely related to static patch holography.

Another way to make a connection with static patch holography in this simple model is by applying generalized T$^2$ deformation in the DSSYK model \cite{Aguilar-Gutierrez:2024oea}. Related ideas have been recently discussed by \cite{Chang:2025ays}, and in upcoming work.

\paragraph{Wormhole Hilbert Space}In this work we have focused on the double-scaled algebra and the chord Hilbert space; particularly for one and zero-particle states. However, there is a different perspective that one could study in the future based on the wormhole Hilbert space construction of sine dilaton gravity in \cite{Cui:2025sgy}. For instance, the factorization structure in their construction with many particles is manifest; given that the wormhole Hilbert space is not isomorphic to the chord Hilbert space, and the exact relation between them has yet to be addressed. Nevertheless, the wormhole Hilbert space reproduces the same results in terms of correlation functions as the chord Hilbert space with matter; {and therefore, it might also reproduce expectation values for other operators in the double-scaled algebra, such as $\hrho\log\hrho$ in the computation of algebraic entanglement entropy \eqref{eq:algebraic entropy}}. It would be interesting to formulate our results in terms of similar expectation values of operators acting on the wormhole Hilbert space, which is based on the end-of-the-world branes in the DSSYK \cite{Okuyama:2023byh,Cao:2025pir,Aguilar-Gutierrez:2025hty,Watanabe:2025rwp}.

\paragraph{Relational Entanglement Entropy with More General QRFs}In Sec.~\ref{sec:RT dS2} we found that the RT formula is relational with respect to different ``clocks'' at the asymptotic boundaries in the effective AdS$_2$ geometry. It would be interesting to further investigate this for other notions of time which depend on the spacetime foliation in the AdS$_2$ space \cite{Parrikar:2025xmz}. To do this, one could develop a notion of bulk time \cite{York:1971hw,York:1972sj,York:1973ia} in sine dilaton gravity, which relies on solving ADM constraints. 
This could have interesting applications to study evolution in closed universes, including the dS$_2$ limit of sine dilaton gravity. We hope to report progress about this in the future.

\paragraph{Bulk Interpretation of the Krylov Complexity Speed Limits}In Sec.~\ref{sec:RT dS2} we identified a bound on the growth of entanglement entropy with time at the semiclassical limit with Krylov spread complexity; while in App.~\ref{app:details speed limits} we include quantum corrections for the speed limits of spread complexity. Given that there have been several works that show there is a bulk interpretation for Krylov state and operator complexity of the DSSYK model, e.g.~\cite{Rabinovici:2023yex,Heller:2024ldz,Aguilar-Gutierrez:2025pqp,Aguilar-Gutierrez:2025mxf,Aguilar-Gutierrez:2025hty}, it would be useful to find a bulk argument for the validity of the speed limit in App.~\ref{app:details speed limits} in terms of wormhole geodesics in AdS$_2$ (in JT gravity or sine dilaton gravity) with $G_N$ corrections.\footnote{It might be useful to connect our results with wormhole velocities recently appearing in \cite{Fu:2025kkh} after calculations in this work were completed.}

\paragraph{Other Limits}In this work, we identified the {IR and} UV triple-scaling limits in the DSSYK Hamiltonian that reproduces the (A)dS$_2$ generators of translations along the asymptotic boundaries {for each spacetime} (as displayed in Fig.~\ref{fig:dS2_particle}). A natural next step is to figure out the corresponding limit for translations along null surfaces in flat space JT gravity \cite{Afshar:2019axx,Afshar:2021qvi} (which is expected to have an interpretation from sine dilaton gravity \cite{Blommaert:2024whf}) in a similar way as in our work but around the middle of the spectrum of the DSSYK; instead of one of the edges. It would be interesting to verify whether the corresponding RT formula, which we conjectured in Sec.~\ref{sec:RT dS2} (including more general cases for sine dilaton gravity) matches with the entanglement entropy computed from the boundary theory in the appropriate triple-scaling limit to get some insights for flat space holography in two-dimensions (which has been associated to the (complex) SYK model in \cite{Afshar:2019axx}. One might also develop an appropriate limit in the eigenstate thermalization hypothesis (ETH) matrix model \cite{Jafferis:2022uhu,Jafferis:2022wez}, which is closely related to the DSSYK but it has finite number of degrees of freedom, similar to those explored by \cite{Okuyama:2025hsd} in the context of dS JT gravity.

\section*{Acknowledgments}
{I appreciate important feedback from Gonçalo Araújo-Regado, Philipp A. Höhn, and Francesco Sartini about the previous version of this work which contributed to formulating the algebraic approach in this version, and for collaboration on related topics.} I thank Takato Mori for initial collaboration and insightful comments; to Andreas Blommaert, David Kolchmeyer, Masamichi Miyaji, Fabio Ori, Yu-ki Suzuki, Thomas Tappeiner, Bilyana Tomova, Erik Tonni, Sandip P. Trivedi, Meng-Ting Wang, Jiuci Xu and Nicolò Zenoni for illuminating discussions; and Takanori Anegawa for pointing out \cite{Jafferis:2019wkd}. I am indebted to Yukawa Institute for Theoretical Physics at Kyoto University, KU Leuven, Perimeter Institute, and Okinawa Institute of Science and Technology where this work was developed during the YITP-I-25-01 ``Black Hole, Quantum Chaos and Quantum Information'', ``The holographic universe'', ``QIQG 2025'' and ``Quantum Reference Frames 2025'' workshops respectively; and again at YITP for the ``Extreme Universe 2025'' YITP-T-
25-01 workshop, where this work was presented and completed. I also thank the High Energy Physics group in UC Santa Barbara for hosting me during initial stages of this manuscript, and for travel support from the QISS consortium, and from YITP in both visits. SEAG is supported by the Okinawa Institute of Science and Technology Graduate University. This project/publication was also made possible through the support of the ID\#62312 grant from the John Templeton Foundation, as part of the ‘The Quantum Information Structure of Spacetime’ Project (QISS), as well as Grant ID\# 62423 from the John Templeton Foundation. The opinions expressed in this project/publication are those of the author(s) and do not necessarily reflect the views of the John Templeton Foundation.

\appendix

\section{Notation}\label{app:notation}
\paragraph{Definitions}
\begin{itemize}[noitemsep]
\item $N$, and $p$: Total number of fermions; and number of all-to-all interactions respectively.
 \item $q:=\rme^{-\lambda}:=\rme^{-\frac{p^2}{2N}}$.
\item $(a;q)_n:=\prod_{k=0}^{n-1}(1-aq^k)$: q-Pochhammer symbol.
\item $(a_1,a_2,\dots a_m;q)_n:=\prod_{i=1}^N(a_i;~q)_n$.
\item $H_n(x|q)$ \eqref{eq:H_n def}: q-Hermite polynomials.
\item $E(\theta)$ \eqref{eq:energy}: Energy spectrum, where $\theta$ is a parametrization.
\item $\mu(\theta)$ \eqref{eq:norm theta}: Energy basis measure.
\item $\hH$, $\hH_{L/R}$ \eqref{eq:energy basis}: zero and one-particle DSSYK Hamiltonians.
\item $\hat{h}\equiv(\sqrt{\lambda}/J)\hH$ \eqref{eq:def h}: Dimensionless Hamiltonian.
\item $:\hat{A}:$ \eqref{eq:normal order}: Normal ordered operator $\hat{A}$, which moves annihilation operators to the left.
\item $\mathcal{H}_m$ \eqref{eq:states notation matter}, $\mathcal{H}_{\rm chord}$ \eqref{eq:H ffull}: Chord Hilbert space with $m$ particle insertions, and its completion with all possible matter insertions.
\item $\ket{\Psi_{{\Delta}}(\tau_L,\tau_R)}$ \eqref{eq:HH state tL tR}, $Z_\Delta(\beta_L,\beta_R)$ \eqref{eq:partition 1 particle}: Two-sided HH state, and the partition function with one-particle insertion.
\item $\ket{\theta}$, $\ket{\Delta;\theta_L,\theta_R}$ \eqref{eq:energy basis}: Energy basis for zero, and one-particle states respectively.
\item $\ket{n}$, $\ket{\Delta;n_L,n_R}$: Chord number basis in the zero and one-particle space
    \item $\ket{\Omega}$: Zero chord number (maximally entangled \cite{Lin:2022rbf})  state.
\item $\hat{n}$, $\hat{P}$: chord number operator and its canonical conjugate.
\item $\mathcal{A}_{L/R}\equiv\qty{\hH_{L/R},\,\hmO_\Delta^{L/R}}$ \eqref{eq:double scaled algebra}: Double-scaled algebras of DSSYK.
\item $k_{\eta}$ \eqref{eq:k eta}: Critical value in the triple-scaled energy spectrum for the WKB approximation. 
\item $\hat{L}_{\rm IR}\equiv\lambda \hat{n}-2\log\lambda$, $\hat{L}_{\rm UV}=\rmi\hat{L}_{\rm IR}$ (\ref{eq:dS length}): Regularized and rescaled chord operator in the UV and IR triple-scaling limits.
\item $\hat{\mathcal{O}}^{L/R}_{\Delta}$ \eqref{eq:reference pets}: Matter chord operator insertion.
\item $\beta(\theta)$ \eqref{eq:beta semiclassical}: Microcanonical physical temperature.
\item $\hat{\rho}$ \eqref{eq:rho from state definition}: Reduced density matrix. 
\item $\hat{\mathcal{F}}_\Delta$ \eqref{eq:isometric factorization}: Isometric factorization map.
\item $\Tr$ \eqref{eq:algebra trace}: Type II algebra trace.
\item $S$ \eqref{eq:algebraic entropy}, $S^{\rm (thermal)}$ \eqref{eq:entropy}, $\mathcal{S}_\alpha$ \eqref{eq:Renyi entropy}, $\Delta S$ \eqref{eq:important entropy difference}: Algebraic entanglement entropy, thermal entropy, Rényi entropy and regularized von Neumann entropy respectively.
\item $\theta_{\rm IR}\equiv\theta$ and $\theta_{\rm UV}\equiv\pi-\theta$ for $\theta_{\rm IR/UV}\ll1$ \eqref{eq:lengths dS2 time2}: Energy parameters in the IR and UV triple-scaling limits respectively.
\item  $\Phi_{\rm (A)dS}$ \eqref{eq:RT surface dS}: Dilaton
$\Phi$ \eqref{eq:Dilaton-gravity theory} in (dS) JT gravity.
\item $r_{\rm BH}$, $r_{\rm CH}$ \eqref{eq:static patch}: Black hole and cosmological horizons radius in (dS) JT gravity respectively.
\item $\gamma$ \eqref{eq:RT surface dS}: RT surface.
\item$\beta_{\rm AdS}$ \eqref{eq:semiclassical (A)dS inv temp}: Fake temperature.
\item $S_0$ \eqref{eq:semiclassical entropy}: Constant term in the DSSYK thermodynamic entropy.
\item $\hH_{\rm IR}$ and $\hH_{\rm UV}$ \eqref{eq:dS Hamiltonian}: DSSYK Hamiltonian in the IR and UV triple-scaling limit \eqref{eq:def IR UV triple scaling limit}.
\item $\mathcal{C}^{(\eta)}$ \eqref{eq:complexity eta}: One-particle state Krylov complexity.
\item $\ket{K_n}$ \eqref{eq:normal ordered total chord number}: Krylov basis.
    \item $a_n$, $b_n$ \eqref{eq:Lanczos coef}: Lanczos coefficients.
\end{itemize}
\paragraph{Acronyms}
\begin{itemize}[noitemsep]
\item (A)dS: (Anti-)de Sitter.
\item ADM: Arnowitt–Deser–Misner
\item BTZ: Bañados-Teitelboim-Zanelli.
\item BH: Bekenstein-Hawking.
\item CA: Complexity=action.
\item CFT: Conformal field theory.
 \item CV: Complexity=volume.
\item (DS)SYK: (Double-scaled) Sachdev-Ye-Kitaev.
\item ETH: Eigenstate thermalization hypothesis.
\item GNS: Gelfand-Naimark-Segal
\item GH: Gibbons-Hawking.
\item HH: Hartle-Hawking.
\item IR: Infrared.
\item JT: Jackiw-Teitelboim.
\item PETS: Partially-entangled thermal states.
\item QRF: Quantum reference frame.
\item RT: Ryu-Takayanagi.
\item WKB: Wentzel-Kramers-Brillouin.
\item UV: Ultraviolet.
\end{itemize}

\section{Sine Dilaton Gravity}\label{app:sine dilaton}
In this appendix we briefly review sine dilaton gravity to keep the manuscript relatively self-contained. This aids the physical interpretation of the results, although it is not necessary to assume {{a correspondence between sine dilaton gravity and}} the DSSYK model \cite{Blommaert:2024ymv} in our computations.

It has been originally argued in \cite{Blommaert:2023opb,Blommaert:2024ymv,Blommaert:2024whf} that the bulk holographic dual of the DSSYK model is a dilaton gravity theory at the disk topology level, which is described in Euclidean-like signature by
\begin{equation}\label{eq:Dilaton-gravity theory}
\begin{aligned}
    I_{\rm SD}=&-\frac{1}{16\pi G_N}\int_{\mathcal{M}}\rmd^2x\sqrt{g}\qty(\Phi\mathcal{R}+2\sin\Phi)+I_{\rm bdry}~,
\end{aligned}
\end{equation}
where $\mathcal{M}$ is the manifold, {{$G_N$ Newton's constant}}, $\mathcal{R}$ the Ricci scalar, $g_{\mu\nu}$ the metric in $\mathcal{M}$, and $I_{\rm bdry}$ the boundary action. The Euclidean-like vacuum solution to the equations of motion of (\ref{eq:Dilaton-gravity theory}) is given by:\footnote{One can add a topological term to the action with $\Phi_0\gg1$ to suppress wormhole topologies \cite{Mertens:2022irh}.}
\begin{equation}\label{eq:metric}
    \rmd s^2=2\qty(\cos r_{\rm BH}-\cos \Phi)\rmd\tau^2+\frac{\rmd \Phi^2}{2\qty(\cos r_{\rm BH}-\cos \Phi)}~,
\end{equation}
where $r_{\rm BH}$ is the black hole horizon, corresponding to $\theta$ \eqref{eq:energy} in the DSSYK model according to \cite{Blommaert:2024ymv}. The effective AdS$_2$ black hole geometry found by Weyl-rescaling the metric (\ref{eq:metric}):
\begin{align}
\label{eq:effective geometry}
    &\rmd s^2_{\rm eff}=\rme^{\mp\rmi \Phi^{(\pm)}}\rmd s^2={F_{\rm eff}(\upsilon)}{\rmd\tau^2}+\frac{\rmd \upsilon^2}{F_{\rm eff}(\upsilon)}~,\quad F_{\rm eff}(\upsilon)=\upsilon^2-\sin^2r_{\rm BH}~,\\
    &\text{where}\quad\Phi^{(\pm)}=\pm\qty(\frac{\pi}{2}+\rmi\log(\upsilon+\rmi\cos r_{\rm BH}))~.\label{eq:change variables}
\end{align}
See \cite{Blommaert:2024ymv} for more details on the effective geometry.

One recovers JT gravity $I_{\rm SD}\rightarrow I_{\rm JT}$ by setting $\Phi\rightarrow \Phi_{\rm AdS}$ ($\abs{\Phi_{\rm AdS}}\ll1$):
\begin{equation}\label{eq:JT-gravity theory}
\begin{aligned}
    I_{\rm JT}=&-\frac{1}{16\pi G_N}\int_{\mathcal{M}}\rmd^2x\sqrt{g}\Phi_{\rm AdS}(\mathcal{R}+2)+I_{\rm bdry}~.
\end{aligned}
\end{equation}
and dS JT gravity by taking $\Phi_{\rm dS}\equiv\pi-\Phi$ with $\abs{\Phi_{\rm dS}}\ll1$ {{at $\mathcal{O}(\Phi_{\rm dS})$ which leads to $I_{\rm SD}\rightarrow I_{\rm dS~JT}$ where}}
\begin{equation}\label{eq:dS JT gravity action}
\begin{aligned}
  I_{\rm dS~JT}=&-\frac{1}{16\pi G_N}\int_{\mathcal{M}}\rmd^2x\sqrt{g}\qty(\Phi_0\mathcal{R}-\Phi_{\rm dS}\qty(\mathcal{R}-2))+I_{\rm bdry}~.
\end{aligned}
\end{equation}
{{Here $\Phi_0\equiv \pi$ so that the topological term above is a finite constant.}} The metric solving the dilaton equation of motion of \eqref{eq:JT-gravity theory} and \eqref{eq:dS JT gravity action} is described by Rindler-AdS$_2$ and the static patch of dS$_2$ in \eqref{eq:static patch} respectively, which can be seen from \eqref{eq:metric} by performing an expansion
\begin{equation}
\begin{aligned}
    {\rm AdS}:\quad&\Phi=\Phi_{\rm dS}=r\ll1~,\quad r_{\rm BH}\ll1~,\\
    {\rm dS}:\quad&\Phi_{\rm dS}=r(\equiv \pi-\Phi)\ll1~,\quad r_{\rm CH}\equiv\pi-r_{\rm BH}\ll1~. 
\end{aligned}
\end{equation}
To describe the rest of the (A)dS$_2$ geometry, one needs to analytically continue the above metric, so that it corresponds to solution of the equations motion of \eqref{eq:JT-gravity theory} and \eqref{eq:dS JT gravity action} without requiring $\abs{\Phi_{\rm (A)dS}}\ll1$ respectively.

{\section{More on the WKB Approximation}\label{app:WKB approx}
In this appendix, we provide additional details about deriving \eqref{eq:k eta} from the algebraic entanglement entropy \eqref{eq:entropy n} using the density matrix associated to the chord number $\hrho=:\hat{h}^n:\ket{\Omega}$ \eqref{eq:rho from n}, where $\hat{h}\equiv\frac{\sqrt{\lambda(1-q)}}{J}\hH$ is the rescaled chord Hamiltonian. 
\\
We are interested in the entropy difference \eqref{eq:important entropy difference}:
\begin{equation}\label{eq:entropy difference}
\begin{aligned}
S&=-\bra{\Omega}\qty(\hrho\log\hrho-\eval{\hrho\log\hrho}_{n\rightarrow\infty})\ket{\Omega}\\
&=-\int_0^\pi\rmd\theta\qty(p_n(\theta)\log \frac{p_n(\theta)}{\mu(\theta)}-{p_{n\rightarrow\infty}(\theta)\log \frac{p_{n\rightarrow\infty}(\theta)}{\mu(\theta)}})~,
\end{aligned}
\end{equation}
where we represent $p_n(\theta)\equiv\mu(\theta)\abs{\bra{\theta}\ket{n}}^2$, with $\mu(\theta)$ in \eqref{eq:norm theta} and $\bra{\theta}\ket{n}$ in \eqref{eq:H_n def}; and $\Tr(\hrho)=1$. To carry out the evaluation \eqref{eq:entropy difference} in the triple-scaling limit, we will specialize in the UV limit \eqref{eq:def IR UV triple scaling limit},
\begin{equation}\label{eq:triple scaling dS again}
   {\rm UV~limit}\quad\rme^{\rmi L_{\rm dS}}\equiv\frac{\rme^{-\lambda n}}{2\lambda^{2}}\quad\text{and}\quad k_{\rm dS}\equiv\frac{\pi-\theta}{\lambda}\quad\text{are fixed as }\lambda\rightarrow0~,
\end{equation}
since it is the focus of Sec.~\ref{sec:RT dS2}. Analogous results apply for the IR case \cite{Tang:2024xgg}, which is the basis of this appendix. \eqref{eq:triple scaling dS again} then leads to
\begin{equation}\label{eq:triple_scaled prob}
p_n(\theta)\equiv\frac{(q,\rme^{\pm2\rmi\theta};q)_\infty}{2\pi}\frac{H_n(\cos\theta|q)}{\sqrt{(q;q)_n}}~\underset{\rm UV}{\rightarrow} ~p(L_{\rm dS},k_{\rm dS})\equiv{\frac{2}{\pi}}\qty(\frac{K_{2\rmi\sqrt{2} k_{\rm dS}}(4\rme^{\rmi L_{\rm dS}/2})}{\Gamma(2\rmi\sqrt{2} k_{\rm dS})})^2~,
\end{equation}
where the arrow denotes the UV triple-scaling limit, and $K_\nu(z)$ is the modified Bessel function of the second kind. Additionally, the probability density \eqref{eq:triple_scaled prob} can be seen as the norm of a wavefunction
\begin{equation}
    \psi_{k_{\rm dS}}(L_{\rm dS})\equiv \sqrt{\frac{2}{\pi}}\frac{K_{2\rmi\sqrt{2}k_{\rm dS}}(4\rme^{\rmi L_{\rm dS}/2})}{\Gamma(2\rmi\sqrt{2} k_{\rm dS})}~,\quad p(L_{\rm dS},k_{\rm dS})=\abs{\psi_{k_{\rm dS}}(L_{\rm dS})}^2~,
\end{equation}
which satisfies the Liouvillian equation resulting from the triple-scaling limit of the zero-particle chord Hamiltonian \eqref{eq:Hamiltonian simpler} with the energy spectrum in \eqref{eq:enerrgy spectrum small}
\begin{equation}\label{eq:EOM triple scaling}
\qty(\frac{1}{2}\partial^2_{L_{\rm dS}}+2\rme^{\rmi L_{\rm dS}})\psi_{k_{\rm dS}}(\ell)={k_{\rm dS}^2}\psi_{k_{\rm dS}}(L_{\rm dS})~,
\end{equation}
where $\rmi\partial_{L_{\rm dS}}$ corresponds to $P_{\rm dS}$ in \eqref{eq:Hamiltonian simpler}. Then, \eqref{eq:EOM triple scaling} can be used to determine the momentum scale where the WKB approximation is valid, corresponding to the vanishing of the kinematic energy in \eqref{eq:EOM triple scaling} (i.e.~$P_{\rm dS}^2=0$), which occurs when \eqref{eq:k eta}
\begin{equation}\label{eq:max momentum}
k_{\eta=-1}\equiv \sqrt{2}~\rme^{\rmi L_{\rm dS}/2}~.
\end{equation}
In the WKB scheme \cite{wentzel1926verallgemeinerung,kramers1926wellenmechanik,brillouin1926mecanique} we consider that the wavefunction in \eqref{eq:EOM triple scaling} can be approximately described by a particle trapped in a potential given by the $2\rme^{\textcolor{orange}{\rmi} L_{\rm dS}}$ term in the Schrödinger equation \eqref{eq:EOM triple scaling} resulting in
\begin{equation}
    \psi_{k_{\rm dS}}(L_{\rm dS})\simeq\begin{cases}
        \frac{1}{\sqrt{2\pi}}\qty({\frac{k_{\rm dS}^2}{2\rme^{\rmi L_{\rm dS}}-k_{\rm dS}^2}})^{1/4}\exp(\rmi\int_{L_{\rm dS}}^{L_0}\rmd L\sqrt{2\rme^{\rmi L}-k_{\rm dS}^2})&k_{\rm dS}\lesssim k_{\eta=-1}~,\\
        \sqrt{\frac{2}{\pi}}\qty({\frac{k_{\rm dS}^2}{k_{\rm dS}^2-2\rme^{\rmi L_{\rm dS}}}})^{1/4}\sin\qty(-\rmi\int^{L_{\rm dS}}\rmd L\sqrt{k^2_{\rm dS}-2\rme^{\rmi L_{\rm dS}}})&k_{\rm dS}>k_{\eta=-1}~,
    \end{cases}
\end{equation}
where $L_0\equiv-\rmi\log (k_{\rm dS}/\sqrt{2})$ (i.e.~the inverse of \eqref{eq:max momentum}) and the proportionality constant is chosen for normalization. By evaluating the integrals above when $\textcolor{orange}{2\rme^{\rmi L_{\rm dS}}}\ll k_{\rm dS}^2$ \textcolor{orange}{for $k_{\rm dS}\lesssim k_{\eta=-1}$, and $\textcolor{orange}{2\rme^{\rmi L_{\rm dS}}}\gg k_{\rm dS}^2$ for $k_{\rm dS}>k_{\eta=-1}$; so that} the probability distribution $p(L_{\rm dS},k_{\rm dS})$ \eqref{eq:triple_scaled prob} at the leading order then becomes
\begin{equation}\label{eq:p L k dS}
p(L_{\rm dS},k_{\rm dS})\sim\begin{cases}
0&k_{\rm dS}\lesssim k_{\eta=-1}~,\\
\frac{2}{\pi}\sin^2(\rmi L_{\rm dS}k_{\rm dS})&k_{\rm dS}>k_{\eta=-1}~.
\end{cases}
\end{equation}
We can now evaluate the entropy difference in \eqref{eq:entropy difference}, using the UV triple-scaled probability density \eqref{eq:triple_scaled prob}
\begin{equation}
\begin{aligned}
&\int_0^\pi \rmd\theta\qty(p_n(\theta)\log p_n(\theta)-p_{n\rightarrow\infty}(\theta)\log p_{n\rightarrow\infty}(\theta))\\
&\underset{\rm UV}{\rightarrow}2\lambda\int_0^\infty\rmd k_{\rm dS}\qty(p(L_{\rm dS},k_{\rm dS})\log p(L_{\rm dS},k_{\rm dS})-p_0\log p_0)=\mathcal{O}(\lambda)~,
\end{aligned}
\end{equation}
where $p_0\equiv\textcolor{orange}{1}/\pi$ corresponds to $p_{n\rightarrow\infty}(\theta)$ in the triple-scaling limit, as seen from \eqref{eq:p L k dS}; and the last equality follows from $p(L_{\rm dS},k_{\rm dS})\sim \mathcal{O}(1)$ in \eqref{eq:triple_scaled prob}.
\\
Thus, the leading order terms contributing to the entropy difference \eqref{eq:entropy difference} are
\begin{equation}\label{eq:intermediate Delta S}
\begin{aligned}
\Delta S\simeq 2\int_0^\infty\rmd k_{\rm dS}\qty(p(L_{\rm dS},k_{\rm dS})-p_0)\lambda\log\mu(\lambda k_{\rm dS})~,
\end{aligned}
\end{equation}
where we use $\mu(\pi-\theta)=\mu(\theta)$ \eqref{eq:norm theta}. We can implement the relation
\begin{equation}\label{eq:lambda mu}
\begin{aligned}
\lambda\log\mu(\theta)\underset{\rm UV}{\rightarrow}-\frac{1}{2}\pi^2+\mathcal{O}(\lambda\log\lambda)~.
\end{aligned}
\end{equation}
Therefore, to leading order in the semiclassical expansion, the entropy difference \eqref{eq:intermediate Delta S} with \textcolor{orange}{\eqref{eq:p L k dS}} becomes
\begin{equation}
\begin{aligned}
\Delta S&\simeq-\pi^2\int_0^\infty\rmd k \qty(p(L_{\rm dS},k_{\rm dS})-\frac{1}{\pi})\textcolor{orange}{\simeq}\pi\int_0^{k_{\eta=-1}}\rmd k_{\rm dS}~.
\end{aligned}
\end{equation}
Thus, we recover the result in the main text \eqref{eq:k eta}.}

\section{Details About the Ryu-Takayanagi Surface in dS\texorpdfstring{$_2$}{}/CFT\texorpdfstring{$_1$}{}}\label{app:details evaluation RTdS2}
In this short appendix we provide more details about how to locate the RT surface points in Sec.~\ref{ssec:EE bulk dS2} from the entangling region points at $\mathcal{I}^\pm$ in dS$_2$; {the analogous AdS$_2$ black hole case has been analyzed in Sec 5.3.~\cite{Tang:2024xgg}}. Consider dS$_2$ space global coordinates
\begin{equation}\label{eq:global coord}
    \rmd s^2=-\rmd T^2+\cosh^2 T~\rmd \varphi^2~.
\end{equation}
Due to the symmetry, when the entangling region is in $\mathcal{I}^+$ we have that $T=0$ corresponds to the RT surface $\gamma$ in \eqref{eq:RT surface dS}; see Fig.~\ref{fig:dS2_particle} (right). Next, we use static path coordinates in \eqref{eq:static patch} from the analytic continuation where \eqref{eq:static patch} describes the Milne patch. The map to global coordinates \eqref{eq:global coord}, results in \footnote{This amounts to $r_{\rm CH}~t\rightarrow r_{\rm CH}~t+\rmi \frac{\pi}{2}$ in the relation between static patch coordinates \eqref{eq:static patch} describing the static patch, instead of the Milne patch; and global coordinates \eqref{eq:global coord} \cite{Faruk:2023uzs}.} 
\begin{subequations}\label{eq:coordinate change}
    \begin{align}
    \sqrt{\frac{r^2}{r_{\rm CH}^2}-1}\cosh(r_{\rm CH} t)&=\sinh T~,\\
    \sqrt{\frac{r^2}{r_{\rm CH}^2}-1}\sinh(r_{\rm CH} t)&=\cos\varphi~\cosh T~.
\end{align}
\end{subequations}
In particular, we focus on the region $\mathcal{I}^+$ where $r\rightarrow\infty$ and $T\rightarrow\infty$ in \eqref{eq:coordinate change}, which means that
\begin{equation}\label{eq:scri plus location}
    \mathcal{I}^+:~\cos\varphi=\tanh(r_{\rm CH}~t)~,
\end{equation}
where $t$ is the static patch time at $T=\infty$.

Now, we seek to evaluate $\Phi_{\rm dS}(\gamma)$, where $\Phi_{\rm dS}=r$ in the static patch. From symmetry in Fig.~\ref{fig:dS2_particle} we can see that $\gamma$ corresponds to $T=0$ when the entangling regions are points at $\mathcal{I}^\pm$. In global coordinates {the dilaton} at the extremal surface corresponds to \eqref{eq:global coord}:
\begin{equation}
    \Phi_{\rm dS}(\gamma)=r_{\rm CH}\sin\varphi~.
\end{equation}
The above result can be expressed with static patch coordinates in the Milne patch (appropriate for the $\mathcal{I}^\pm$) using the analytic continuation below \eqref{eq:coordinate change} as
\begin{equation}\label{eq:PhidS}
    \Phi_{\rm dS}(\gamma)=r_{\rm CH}\sech(r_{\rm CH} t)~.
\end{equation}
Using \eqref{eq:PhidS} and \eqref{eq:lengths dS2 time} we then recover \eqref{eq:dilaton dS terms}.

\section{Relational Entanglement Entropy}\label{app:relational EE}
In this appendix, we study the interpretation of the algebraic entanglement entropy in (A)dS$_2$/CFT$_1$ \eqref{eq:dilaton dS terms} as relational entanglement entropy \cite{DeVuyst:2024pop,DeVuyst:2024uvd,Araujo-Regado:2025ejs} in terms of QRFs (see e.g.~\cite{DeVuyst:2025ezt,Araujo-Regado:2025ejs,Araujo-Regado:2024dpr,Krumm:2020fws,Hohn:2017cpr,Hoehn:2019fsy,Hoehn:2020epv,Hoehn:2023ehz,Vanrietvelde:2018pgb,Vanrietvelde:2018dit,delaHamette:2021oex,Hohn:2018toe,Hohn:2018iwn,Hoehn:2021flk,Giacomini:2021gei,Yang_2020,DeVuyst:2024uvd,DeVuyst:2024pop,AliAhmad:2024qrf,AliAhmad:2024vdw,AliAhmad:2024wja,DeVuyst:2025ezt} among others) from the bulk perspective.

Considering the sine dilaton gravity interpretation, we introduce the QRF formalism (see e.g.~\cite{Hoehn:2019fsy,delaHamette:2021oex,Hoehn:2023ehz}) to describe clock-like observers at the asymptotic boundary of the bulk theory, corresponding to $\mathcal{I}^\pm$ in the dS$_2$ geometry, Fig.~\ref{fig:dS2_particle}, and the time-like boundaries in the effective AdS$_2$ geometry in Fig.~\ref{fig:bulk_picture}.\footnote{Related discussions about QRFs in the {bulk dual of the DSSYK model} have appeared in \cite{Blommaert:2025eps,Aguilar-Gutierrez:2025hty,Aguilar-Gutierrez:2025sqh}.} The chord Hilbert space (isometric to the dual bulk physical Hilbert space) is embedded inside the bulk kinematical Hilbert space describing a tensor product between the bulk interior of the spacetime (which we denote $\mathcal{H}_{\rm S}$, system) and its asymptotic boundary (which we denote $\mathcal{H}_{\rm C}$, clock)
\begin{equation}
    \mathcal{H}_{\rm phys}\in\mathcal{H}_{\rm kin}\equiv \mathcal{H}_{\rm S}\otimes\mathcal{H}_{\rm C}~.
\end{equation}
where $\mathcal{H}_{\rm phys}$ is isometric to $\mathcal{H}_{1}$ \eqref{eq:states notation matter} (for $m=1$), which can be extended with matter insertions \cite{Lin:2022rbf}. In the kinematical Hilbert space, the Wheeler-DeWitt constraint, as well as physical constraints, such as momentum shift symmetry in sine dilaton gravity \cite{Blommaert:2024whf}, are not imposed. The construction of the system Hilbert space in sine dilaton gravity and the Page-Wootters reduction follows analogously to the supersymmetric case studied in \cite{Aguilar-Gutierrez:2025sqh}. We will focus on the observer Hilbert space.

We define the observer state at the asymptotic boundary through a clock-like label as $\ket{t}$, so that the Hilbert space is defined as
\begin{equation}\label{eq:system Hilbert space}
    \mathcal{H}_C={\rm span}\qty{\ket{t}~|t\in\mathbb{R}}~,\quad \ket{t}=\rme^{-\rmi \hH_{\rm (A)dS} t}\ket{\Omega}~,
\end{equation}
where $\ket{\Omega}$ is the isometric dual to the zero chord state in $\mathcal{H}_0$ of the chord Hilbert space (\eqref{eq:states notation matter} for $m=0$) in the physical bulk Hilbert space, representing a zero temperature maximally mixed state \cite{Lin:2022rbf}; $\hH_{\rm (A)dS}$ are the generators of translations along the asymptotic boundaries in (A)dS$_2$ space dual to the IR/UV limits of the chord Hamiltonian in \eqref{eq:dS Hamiltonian}. Note however, that the static patch time in \eqref{eq:system Hilbert space} defined by analytic continuation in the Milne patch for the dS$_2$ geometry, instead of the static patch itself. While in most, if not all, of the QRF literature on dS space (e.g.~\cite{DeVuyst:2024pop,Fewster:2024pur,DeVuyst:2024uvd,AliAhmad:2024qrf}), the QRFs are defined with respect to a worldline observer \cite{Witten:2023qsv,Witten:2023xze,Chandrasekaran:2022cip}; here we provide a new interpretation of observers at $\mathcal{I}^\pm$ for dS/CFT. This correspond to asymptotic boundary ones in the effective AdS$_2$ geometry in sine dilaton gravity \cite{Blommaert:2024whf}. 
The observer clock time orientations \cite{Hoehn:2019fsy} are defined by translation on the states $\rme^{-\rmi\hH_{\rm (A)dS}t}\ket{t'}=\ket{t'+t}$ which correspond to a physical transformation of the QRFs \cite{Hoehn:2019fsy,Hoehn:2023ehz,delaHamette:2021oex}. 
Then, we interpret the result in \eqref{eq:dilaton dS terms} in terms of relational entanglement entropy \cite{Araujo-Regado:2025ejs,DeVuyst:2024pop,DeVuyst:2024uvd} for different time or space-like separated points at the asymptotic boundaries {in the AdS and dS spaces respectively}. As found in Sec.~\ref{ssec: BH GH formulas}, the entanglement entropy at $t=0$ corresponds to the {BH and} GH entropy \eqref{eq:GH term}, while other locations at the asymptotic boundaries for $t\neq0$ lead to smaller entanglement entropies (as seen from (\ref{eq:dilaton dS terms}, \ref{eq:lengths dS2 time})). This agrees with the existence of a maximal mixed state in type II$_1$ von Neumann algebras which describe the dS$_2$ in the semiclassical limit and DSSYK model $\forall q\in[0,1)$ \cite{Xu:2024hoc,Lin:2022rbf,Tang:2024xgg}{; while for the AdS$_2$ black hole case, we expect the result is modified once we introduce matter in the computation given that the algebra is type II$_\infty$ \cite{Witten:2021unn}. Both cases} display the relational nature of entanglement entropies, pointed out in \cite{DeVuyst:2024pop,DeVuyst:2024uvd,Araujo-Regado:2025ejs}, given that different spatially separated entangling surfaces (identified as QRFs) at the asymptotic boundaries measure different holographic entanglement entropies, as found in \eqref{eq:dilaton dS terms}.

\section{Bounds on Krylov Complexity Growth in Chord Space}\label{app:details speed limits}
In this appendix, we complement our discussion in Sec.~\ref{ssec:Lloyd} by studying bounds on the rate of growth of Krylov complexity in the DSSYK model. Throughout the appendix, we specialize the results for the case of Krylov complexity with the HH state with one operator insertion or no operators as a reference state. The case with a particle insertion allows to study both Krylov operator and state complexity at once, as we summarize below.

\paragraph{Outline}In App.~\ref{sapp:Lanzos} we quickly summarize previous results Krylov complexity in the DSSYK with matter chord operators. In App.~\ref{sapp:lloyd}, we then introduce a Lloyd bound for Krylov spread complexity using the HH state as reference. In App.~\ref{sapp:speed} we study limits on the rate of growth of Krylov complexity.

\subsection{Krylov Complexity with One-Particle Insertion}\label{sapp:Lanzos}
We briefly review some of the results in our previous work \cite{Aguilar-Gutierrez:2025pqp} regarding the construction of the Krylov basis for the two-sided HH state as a reference state.

We begin with the one-particle HH state \eqref{eq:HH state tL tR} expressed as
\begin{equation}\label{eq:Lanczos eta}
    \ket{\Psi_{\Delta}(\tau_L=\eta\tau,\tau_R=\tau)}=\rme^{-\hat{\mathcal{L}}_\eta\tau}\ket{\Delta;0,0}=\sum_n\Psi_n^{(\eta)}(\tau)\ket{K^{(\eta)}_n}~,
\end{equation}
where $\hat{\mathcal{L}}_\eta\equiv\hH_R+\eta\hH_L$. Meanwhile $\qty{\ket{K_n^{(\eta)}}}$ are the Krylov basis for the reference state $\ket{K_0^{(\eta)}}=\ket{\Delta;0,0}$, which satisfy a Lanczos algorithm
\begin{equation}
    \hat{\mathcal{L}}_\eta\ket{K^{(\eta)}_n}=a^{(\eta)}\ket{K^{(\eta)}_n}+b^{(\eta)}_{n+1}\ket{K_{n+1}^{(\eta)}}+b^{(\eta)}_{n}\ket{K_{n-1}^{(\eta)}}~,
\end{equation}
with $b_0^{(\eta)}=0$. The Lanczos algorithm is solved by
\begin{equation}\label{eq:normal ordered total chord number}
    \ket{K^{(\eta)}_n}=c^{(\eta)}_n\sum_{l=0}^n\eta^k\begin{pmatrix}
        n\\
        l
    \end{pmatrix}\ket{\Delta;l,n-l}+\text{additional~terms}~,
\end{equation}
with $c^{(\eta)}_n=\sqrt{\frac{\lambda^n(1-q)^n(1+\eta)}{2^n(q^{1/2};q^{1/2})_n(-\eta;q^{1/2})_{n+1}}}$, while the norm of the additional terms vanishes as $\lambda\rightarrow0$ \cite{Ambrosini:2024sre}. The corresponding \emph{Lanczos coefficients} are
\begin{equation}\label{eq:Lanczos coef}
a^{(\eta)}_n=0~,\quad b^{(\eta)}_n\eqlambda\frac{-2J}{\sqrt{\lambda(1-q)}}\sqrt{(1-q^{n/2})(1+\eta q^{n/2+\Delta})}~.    
\end{equation}
We define Krylov complexity operator for states and operators (i.e.~$\eta=\pm1$) and its generating function respectively as \cite{Aguilar-Gutierrez:2025pqp}
\begin{subequations}\label{eq:relevant defs}
\begin{align}
    &\hat{\mathcal{C}}\equiv \sum_nn\ket{K_n^{(\eta)}}\bra{K_n^{(\eta)}}~,\label{eq:krylov op}\\
    &\mathcal{G}_{\Delta}^{(\Delta_w)}(\tau_L,\tau_R)\equiv\frac{\bra{\Psi_{\Delta}(\tau_L,\tau_R)}q^{\Delta_w\hat{\mathcal{C}}}\ket{\Psi_{\Delta}(\tau_L,\tau_R)}}{Z_{\Delta}(\beta_L,\beta_R)}~,
\end{align}
\end{subequations}
such that the Krylov complexity is recovered by taking expectation values of the Krylov complexity operator on the state used to construct the Krylov basis $\ket{K_n^{(\eta)}}$:
\begin{equation}\label{eq:complexity eta}
    \mathcal{C}^{(\eta)}(t)\equiv \eval{\frac{\bra{\Psi_{\Delta}(\eta\tau,\tau)}\hat{\mathcal{C}}\ket{\Psi_{\Delta}(\eta\tau,\tau)}}{\bra{\Psi_{\Delta}(\eta\tau,\tau)}\ket{\Psi_{\Delta}(\eta\tau,\tau)}}}_{\tau=\frac{\beta}{2}+\rmi t}~,
\end{equation}
and the Krylov complexity as well as higher moments can be calculated by taking derivatives of the generating function. Note that $\hat{\mathcal{C}}$ does not necessarily equal the total chord number operator in the DSSYK; however, it does in the semiclassical limit \cite{Aguilar-Gutierrez:2025pqp}. The Krylov complexity of states and operators are \cite{Aguilar-Gutierrez:2025pqp}
\begin{subequations}\label{eq:complexity eta pm1}
    \begin{align}\label{eq:Krylov complexity many}
        &\mathcal{C}^{(\eta=\pm1)}(t)\eqlambda\frac{2}{\lambda}\log(A(\theta, \pm q^{\Delta})+B(\theta, \pm q^{\Delta})\cosh(2J\sin\theta t))~,\\
&A(\theta,q^\Delta)=\frac{(\cos \theta
   -\cos 2 \theta ) q^{\Delta
   }+\left(1-q^{\Delta }\right) \left(q^{\Delta
   }+\sqrt{q^{2 \Delta }-2 \cos \theta 
   q^{\Delta }+1}-1\right)}{2
   \sin ^2\theta \left(q^{\Delta }+\sqrt{q^{2 \Delta }-2 \cos
   \theta  q^{\Delta }+1}-1\right)}~,\\
&B(\theta,q^\Delta)=\frac{q^{2 \Delta
   }-\left((\cos \theta +\cos 2 \theta )
   q^{\Delta }\right)-\left(1-q^{\Delta
   }\right) \sqrt{q^{2 \Delta }-2 \cos \theta
    q^{\Delta }+1}+1}{2
   \sin ^2\theta \left(q^{\Delta }+\sqrt{q^{2 \Delta }-2 \cos
   \theta  q^{\Delta }+1}-1\right)}~.\label{eq:womrhole semiclassical answer many}
    \end{align}
\end{subequations}

\subsection{Lloyd Bound}\label{sapp:lloyd}
In this subsection we identify a Lloyd bound for Krylov complexity in the DSSYK model with matter operator insertions. We are motivated by recent results suggesting that Krylov complexity in the DSSYK model \cite{Rabinovici:2023yex,Ambrosini:2024sre,Ambrosini:2025hvo,Heller:2024ldz,Heller:2025ddj,Aguilar-Gutierrez:2025pqp,Aguilar-Gutierrez:2025hty,Aguilar-Gutierrez:2025sqh,Aguilar-Gutierrez:2025mxf} is an explicit realization of the complexity=volume (CV) proposal \cite{Susskind:2014rva,Stanford:2014jda}. Originally, the {Lloyd bound}~\cite{lloyd2000ultimate} was introduced as
information processing rate bound in general physical devices. It was later adapted for the CV and complexity=action (CA) proposals in holographic complexity by \cite{Brown:2015bva,Brown:2015lvg} for asymptotically AdS spacetimes black holes. According to the proposal, the maximum rate of growth of holographic complexity is proportional to a combination of thermodynamic quantities for the system under consideration.\footnote{The conditions under which the Lloyd bound in the CA proposal is satisfied have been studied in detail in \cite{Yang:2016awy}, while different violations of the Lloyd bound in the CA proposal have been reported in \cite{Carmi:2017jqz, Yang:2017czx,Bernamonti:2021jyu,Auzzi:2018pbc,Couch:2017yil,Swingle:2017zcd, An:2018xhv, Alishahiha:2018tep, Wang:2023ipy}. Violations of the Lloyd bound in the CV proposal are rare in comparison with the CA proposal  \cite{Auzzi:2023qbm,Zolfi:2023bdp,Ageev1,Ageev2,Aguilar-Gutierrez:2023ccv}. The conditions under which the CV proposal in \cite{Engelhardt:2021mju} holds have been rigorously investigated in asymptotically AdS$_{d+1\geq4}$ spacetimes with minimally coupled Maxwell-scalar matter that obeys the weak energy condition \cite{Engelhardt:2021mju}.}

The Krylov spread complexity for the zero-particle HH state, which we will denote $\mathcal{C}(t)$, is given by the expectation value of the chord number in the HH state $n(t)$ \eqref{eq:n HH state} in the main text; or equivalently \eqref{eq:Krylov complexity many} for $\eta=+1$ and $\Delta=0$. We are interested in the late time limit, where spread complexity saturates to a constant, namely:
\begin{equation}\label{eq:lloyd Krylov}
    \lim_{t\rightarrow\infty}\dv{t}\mathcal{C}=\frac{2}{\lambda}J\sin\theta+\mathcal{O}(\lambda)\propto \frac{S_0}{\beta_{\rm fake}}~,\quad \beta_{\rm fake}=\frac{2\pi}{\sin\theta}
\end{equation}
where $S_0$ is the leading proportionality constant in the thermodynamic entropy of the DSSYK model (see \eqref{eq:semiclassical entropy}) $S=S_0+\mathcal{O}(\frac{1}{\lambda})$; and $\beta_{\rm fake}$ is the ``fake'' inverse temperature of the DSSYK \cite{Lin:2023trc}. 

Note that while there is a significant one-loop correction to the thermodynamic entropy in the semiclassical limit (displayed in \eqref{eq:semiclassical entropy}); $S_0$ is proportional to the number of Majorana fermions \cite{Goel:2023svz}, which is infinite in the DSSYK model. This implies that we can approximate the entropy by the numerical constant $S_0$. Therefore, we observe that \eqref{eq:lloyd Krylov} indeed has the appropriate structure of the Lloyd bound in holographic complexity \cite{Brown:2015bva,Brown:2015lvg}, where the derivative of a complexity measure is proportional to the product of the thermodynamic entropy and the temperature of the system.\footnote{Another way to express the Lloyd bound is in terms of the thermodynamic energy \cite{Das:2024zuu}; however, \eqref{eq:lloyd Krylov} is not be manifestly related to the DSSYK energy spectrum \eqref{eq:energy basis}.} Thus, at leading order in the semiclassical there is a Lloyd bound in Krylov spread complexity \eqref{eq:lloyd Krylov}; which also determines the growth of the algebraic entanglement entropy for chord number states in DSSYK as an exponential function of the Krylov spread complexity of the HH state according to \eqref{eq:speed bound dS HEE}.

Below, we incorporate leading order quantum corrections to study the bounds on the spread complexity rate of growth. However, once we incorporate first order quantum correction in the semiclassical length, it might not satisfy a simple relation with the holographic entanglement  entropy derived in \eqref{eq:entropy Delta dS2}.

\subsection{Speed Limits}\label{sapp:speed}
Now, we incorporate quantum fluctuations in the bound using the speed limits based on the Robertson uncertainty relation \cite{robertson1929uncertainty}, first defined \cite{Hornedal:2022pkc} for Krylov operator complexity (which can be used more universally). This provides a sharper analogue of the Lloyd bound for Krylov complexity above, since one can include quantum corrections. 

In general, the Robertson uncertainty relation states that \cite{robertson1929uncertainty}
\begin{equation}\label{eq:Robertson UR}
    \abs{\expval{[\hat{A}_1,\hat{A}_2]}}\leq 2\sqrt{{\rm Var}(\hat{A}_1){\rm Var}(\hat{A}_2)}~,
\end{equation}
where ${\rm Var}(\hat{A}_i)\equiv {\expval{\hat{A}_i^2}-\expval{\hat{A}_i}^2}$, and the expectation values are taken with respect to an arbitrary state. We would like to study the consequences of (\ref{eq:Robertson UR}) by the operators $\hat{A}_{1,2}$ to be the Krylov complexity operator $\hat{\mathcal{C}}$ (\ref{eq:krylov op}) and being the generalized Liouvillian operator $\hat{\mathcal{L}}_\eta\equiv \hH_R+\eta\hH_L$ in App.~\ref{sapp:Lanzos}; as well as to specialize in the two-sided HH state $\ket{\Psi_{\Delta}(\eta\tau,\tau)}=\rme^{-\tau\hat{\mathcal{L}}_\eta}\ket{\Delta;0,0}$ (\ref{eq:HH state tL tR}) with $\tau=\frac{\beta}{2}+\rmi t$ to take expectation values. We first notice that:
\begin{equation}\label{eq:var Liouvillian}
\begin{aligned}
    {\rm Var}(\hat{\mathcal{L}}_\eta)&\equiv {\bra{\Delta_S;0,0}\hat{\mathcal{L}}_\eta^2\rme^{-\beta \hat{\mathcal{L}}_\eta}\ket{\Delta_S;0,0}-\qty(\bra{\Delta_S;0,0}\hat{\mathcal{L}}_\eta\rme^{-\beta \hat{\mathcal{L}}_\eta}\ket{\Delta_S;0,0})^2}\\
    &={\dv[2]{Z_{\Delta}(\eta\beta,\beta)}{\beta}-\qty(\dv{Z_{\Delta}(\eta\beta,\beta)}{\beta})^2}~,
\end{aligned}
\end{equation}
where in the first line we used the fact that $\hat{\mathcal{L}}_\eta$ is Hermitian, and in the second one we used the definition of the one-particle partition function $Z_{\Delta}$ \eqref{eq:partition 1 particle}. In particular, when we consider the infinite temperature limit in the first line of (\ref{eq:var Liouvillian}) we can recover (and even extend) the result by \cite{Hornedal:2022pkc}. Namely 
\begin{equation}
    \sqrt{{\rm Var}(\hat{\mathcal{L}}_\eta)_{\beta=0}}=\sqrt{\abs{\bra{\Delta;0,0}\hat{\mathcal{L}}_\eta^2\ket{\Delta;0,0}}}=b^{(\eta)}_1~,
\end{equation}
where $b^{(\eta)}_n$ appears in \eqref{eq:Lanczos coef}, so that the relation above leads to a similar speed limit as \cite{Hornedal:2022pkc} in Krylov operator complexity (for $\eta=-1$). In particular, it also applies for spread complexity ($\eta=+1$). Meanwhile, for  general $\eta\in\mathbb{R}$, these expressions correspond to extended notions of Krylov complexity in for two-sided Hamiltonians \cite{Aguilar-Gutierrez:2025pqp}.

Next, the standard deviation of the Krylov complexity operator (\ref{eq:Krylov complexity many}) gives 
\begin{equation}
\begin{aligned}\label{eq:variance Krylov}
    &{\rm Var}(\hat{\mathcal{C}})={\eval{\dv{\mathcal{G}_{\Delta}^{(\Delta_w)}(\tau_L=\eta\tau,\tau_R=\tau)}{\Delta_w}}_{\Delta_w=0}-\qty(\mathcal{C}^{(\eta)}(t))^2}~,
\end{aligned}
\end{equation}
where the above quantities are defined in \eqref{eq:relevant defs}. Note that the Krylov complexity generating function above does not need to agree with the chord number generating function (i.e.~the thermal two-sided two-point function), which we denote $G_{\Delta}^{(\Delta_w)}(\tau_L,\tau_R)$, although they do agree in the semiclassical limit \cite{Aguilar-Gutierrez:2025pqp}.

On the other hand, the commutator in (\ref{eq:Robertson UR}) for $\hat{\mathcal{L}}_\eta$ and $\hat{\mathcal{C}}$ in the state (\ref{eq:HH state tL tR}) gives
\begin{equation}\label{eq:new Robertson}
    \abs{\dv{t}\mathcal{C}^{(\eta)}(t)}\leq\sqrt{{\rm Var}(\hat{\mathcal{L}}_\eta){\rm Var}(\hat{\mathcal{C}})}~,
\end{equation}
{where we performed a total derivative with respect to $t_R=\eta t_L=:t$. Notice however that for single-sided Hamiltonian systems, we take the derivative with respect to a single coordinate time $t$ resulting in an additional factor $2$ in the left-hand side of (\ref{eq:new Robertson}).}

So far, we derived extensions of the original bound \cite{Hornedal:2022pkc} for two-sided Hamiltonians considering our definition of the two-sided HH state for taking expectation values. However, we can see that the classical Krylov generating function \cite{Aguilar-Gutierrez:2025pqp} which is obtained from the replacement of operator expectation values for classical variables, i.e.~$\expval{\hat{\mathcal{C}}^k}=\expval{\hat{\mathcal{C}}}^k$ (see more details in \cite{Aguilar-Gutierrez:2025pqp}), would violate the Robertson bound. One can easily confirm this noticing that the left-hand side of (\ref{eq:new Robertson}) is non-vanishing generically (using for instance (\ref{eq:womrhole semiclassical answer many}) for $\partial_{t}\mathcal{C}^{(\eta)}(t)$). One needs at least first order quantum corrections to confirm the validity of the bound. This is the point we address next.

\paragraph{Zero-Particle Space Analysis}In order to explicitly confirm the new bound problem (\ref{eq:new Robertson}), we will focus our discussion on Krylov state complexity for the HH state in the $\mathcal{H}_{0}$. This means, $\Delta=0$ and $\eta=+1$ in the previous expressions. In particular, the HH state can be represented as in \eqref{eq:HH state}. Given that we consider a single-sided Hamiltonian system, the speed limit now becomes (dropping the $\eta$ index from now on),
\begin{equation}\label{eq:Robertson B single sided}
    \abs{\dv{t}\mathcal{C}(t)}\leq2\sqrt{\qty(\dv[2]{Z}{\beta}-\qty(\dv{Z}{\beta})^2){\rm Var}(\hat{\mathcal{C}})}~,
\end{equation}
where the term in parenthesis follows from (\ref{eq:var Liouvillian}) and we use the partition function in $\mathcal{H}_0$, $Z(\beta)=\bra{\Omega}\rme^{-\beta\hH}\ket{\Omega}$. For completeness, we present the partition function including the one-loop corrections \cite{Goel:2023svz}
\begin{equation}
   Z(\beta)=\exp[-\frac{2}{\lambda}\qty(\qty(\frac{\pi}{2}-\theta)^2-\qty(\pi-2\theta)\cot\theta)]\frac{\sin\theta~\rme^{(\frac{\pi}{2}-\theta)\cot\theta}}{\sqrt{1+\qty(\frac{\pi}{2}-\theta)\cot\theta}}~.
\end{equation}
However, since we are interested in the leading order analysis in $\lambda$ of the Robertson bound for spread complexity in (\ref{eq:Robertson B single sided}), the one-loop corrections only play an important role for Var$(\hat{\mathcal{C}})$. 

\begin{figure}
    \centering
    \includegraphics[width=0.65\textwidth]{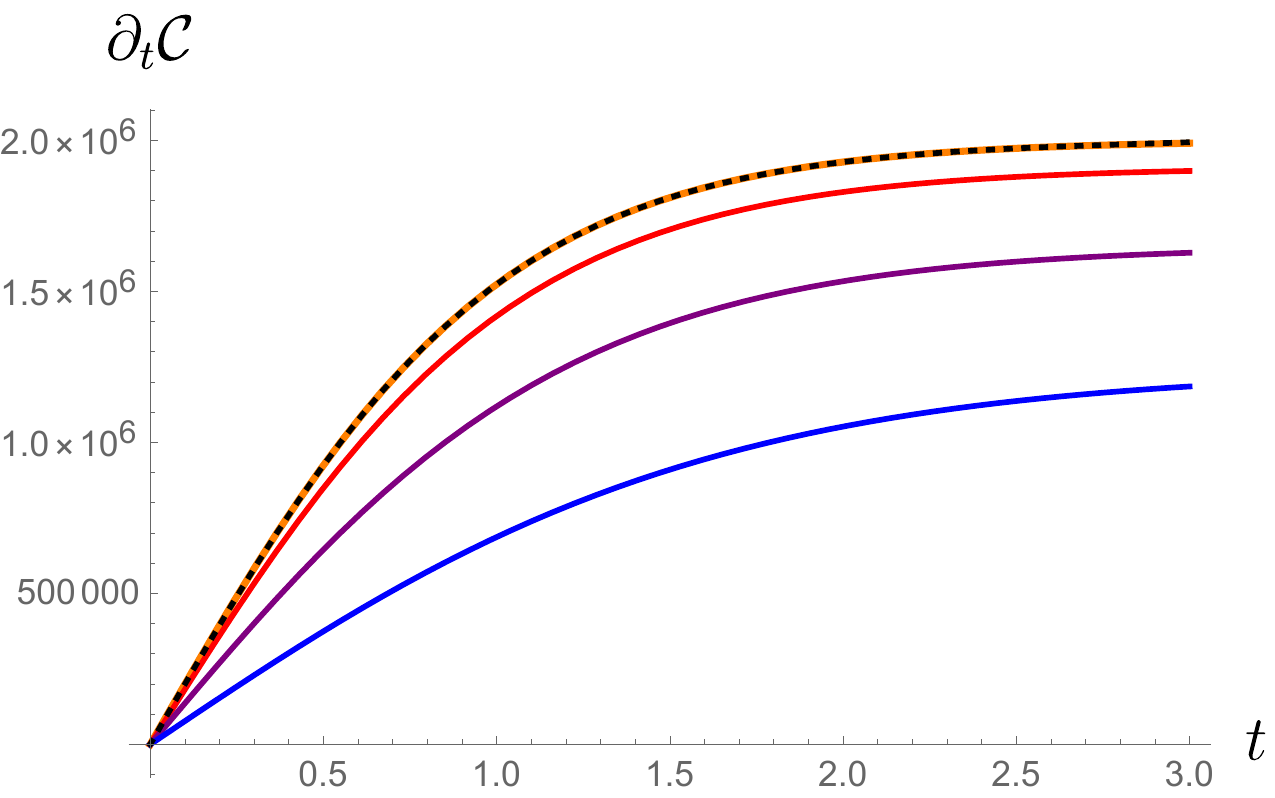}
    \caption{Rate of growth of spread complexity of the zero-particle HH state with one-loop quantum corrections (solid lines) vs the speed limit in Krylov complexity (\ref{eq:new Robertson DSSYK}) for $\theta=\pi/2$ (dotted black line). The orange solid curve corresponds to $\theta=\pi/2$, while the solid lines represent a decrease in multiples of $-0.3$ from $\theta=\pi/2$ in descending order. Similar results are obtained for other values of $\theta$. In all cases the bound (\ref{eq:new Robertson DSSYK}) is satisfied.}
    \label{fig:speed bounds}
\end{figure}

We proceed by evaluating the first and second Krylov moments in (\ref{eq:variance Krylov}) from the generating function, as a series in $\lambda$ \cite{Goel:2023svz}
\begin{equation}\label{eq:generating function 0 particle}
\begin{aligned}
    G^{(\Delta)}(\tau)&=\frac{\bra{\Psi(\tau)}q^{\Delta\hat{n}}\ket{\Psi(\tau)}}{\bra{\Omega}\rme^{-\beta\hH}\ket{\Omega}}\\
    &=\qty(\frac{\sin\theta}{\cosh\qty(J\sin\theta~t)})^{2\Delta}\qty(1+\lambda\qty(\Delta^2\mathcal{I}+\Delta~\mathcal{A})+\mathcal{O}(\lambda^2))~.
\end{aligned}
\end{equation}
Labeling $w=\pi/2-\theta$, and $z=\pi/2-\theta-J\sin\theta\qty(\frac{\beta(\theta)}{2}-\rmi t)$, we have \cite{Goel:2023svz}
\begin{align}\label{eq:correction I}
    \mathcal{I}&=-\frac{((w+z) \tan w \tan z+\tan w+\tan z) (\tan w
   ((w-z) \tan z-1)+\tan z)}{w \tan w+1}~,\\
   \mathcal{A}&=\frac{z^{2 \left(\tan ^2w-\tan ^2z\right)-\frac{z \tan
   z+1}{w \tan w+1}+1}-(w \tan w+1)^2 \sec ^2z+\sec
   ^2w (z \tan z+1)^2}{2 (w \tan w+1)}~.\label{eq:correction A}
\end{align}
The correction in the generating function also allows us to evaluate the first order quantum correction to spread complexity in the DSSYK model without matter, namely
\begin{equation}\label{eq:correction 2pnt}
    \mathcal{C}(t)=\frac{2}{\lambda}\log\frac{\cosh\qty(J\sin\theta~t)}{\sin\theta}\qty(1-\lambda\mathcal{A}(t)+\mathcal{O}(\lambda^2))~.
\end{equation}
The speed limit in (\ref{eq:Robertson B single sided}) can be then expressed as
\begin{equation}\label{eq:new Robertson DSSYK}
\begin{aligned}
    \abs{\partial_t \mathcal{C}}\leq  &\sqrt{\frac{2 \left(4-(\lambda -4)
   \cos (2 \theta )+4 (\pi -2 \theta )
   \cos ^2\theta  \cot \theta
   +\lambda \right)}{2(\pi -2 \theta )
   \cot \theta +4}-4 \cos ^2\theta}~\cdot\\
   &\cdot\frac{4J\sqrt{-2\mathcal{I}}}{\lambda^{3/2}}\rme^{\frac{(\pi -2 \theta ) (2
   \theta +4 \cot (\theta )-\pi
   )}{2\lambda }}+\mathcal{O}(\lambda^{-1})~.
   \end{aligned}
\end{equation}
Therefore, employing (\ref{eq:generating function 0 particle}), we evaluated the rate of growth of spread complexity and compared it to the speed limit (\ref{eq:new Robertson DSSYK}). The results are shown in Fig.~\ref{fig:speed bounds}.

In short, in all cases we evaluate the speed limit (\ref{eq:new Robertson}) (with $\Delta=0$, $\eta=1$) is always satisfied. In particular, when $\theta=\pi/2$, the bound is saturated by $\partial_t\mathcal{C}(t)$ at least for the one-loop quantum corrected Krylov complexity. Meanwhile for other values of $\theta$, there is no saturation, instead the bound is generically much larger than $\partial_t\mathcal{C}(t)$ itself. 

One may also repeat our analysis in the $\mathcal{H}_1$ irrep. while including quantum corrections to evaluate the first and second Krylov complexity moments in ${\rm Var}(\hat{\mathcal{C}})$. A technical complication is that the Krylov basis is no longer approximated by the analytic ansatz in (\ref{eq:normal ordered total chord number}), as discussed in \cite{Ambrosini:2024sre,Aguilar-Gutierrez:2025pqp}. This means that one needs to construct the Krylov basis explicitly numerically. Nevertheless, we expect that the speed limit (\ref{eq:new Robertson}) will be obeyed again. 
\bibliographystyle{JHEP}
\bibliography{references.bib}
\end{document}